\documentclass[12pt,a4]{article}
\usepackage{amsthm,amsmath,latexsym,amssymb,amsfonts,amscd}
\usepackage{graphics,lscape,fancyhdr,array,stmaryrd,euscript,bbm}
\usepackage{ifthen,empheq,slashed}
\usepackage{ifpdf}
\usepackage{verbatim}
\numberwithin{equation}{section}
\usepackage{tikz}
\usepackage{hyperref}

\usepackage[numbers,sort&compress]{natbib}
\usepackage[nottoc]{tocbibind}

\newcommand{\pl}{\partial}

\newcommand{\be}{\begin{equation}}
\newcommand{\ee}{\end{equation}}
\newcommand{\bea}{\begin{eqnarray}}
\newcommand{\eea}{\end{eqnarray}}

\newcommand{\fud}[2]{{}^{#1}{}_{#2}\,}

\newcommand{\besubeqs}{\begin{subequations}}
\newcommand{\esubeqs}{\end{subequations}}

\newcommand{\hsundeformed}{\mathcal{A}}
\newcommand{\hsdeformed}{{\mathcal{A}_\nu}}
\DeclareMathOperator{\sign}{sign}
\newcommand{\Tr}{\mathrm{Tr}}
\newcommand{\omegab}{{\boldsymbol{\omega}}}
\newcommand{\Cb}{{\boldsymbol{C}}}
\newcommand{\Rb}{{\boldsymbol{R}}}

\newcommand{\omegabb}{{\boldsymbol{\hat{\omega}}}}
\newcommand{\Cbb}{{\boldsymbol{\hat{C}}}}

\newcommand{\gbb}{{\boldsymbol{\hat{g}}}}

\begin{document}
\pagenumbering{gobble}
\hfill
\vskip 0.02\textheight
\begin{center}
{\large\bfseries 
Higher-Spin Poisson Sigma Models\\ and\\\vspace{2mm} Holographic Duality for SYK Models}

\vspace{0.4cm}

\vskip 0.03\textheight
\renewcommand{\thefootnote}{\fnsymbol{footnote}}
Xavier \textsc{Bekaert}${}^{a}$,
Alexey \textsc{Sharapov}${}^{b}$ \&
Evgeny \textsc{Skvortsov}\footnote{Research Associate of the Fund for Scientific Research -- FNRS, Belgium.}${}^{c}$\footnote{Also affiliated with Lebedev Institute of Physics.} 
\renewcommand{\thefootnote}{\arabic{footnote}}
\vskip 0.03\textheight

{\em ${}^{a}$Institut Denis Poisson, Unit\'e Mixte de Recherche $7013$ du CNRS,\\ 
Universit\'e de Tours \& Universit\'e d'Orl\'eans,\\
Parc de Grandmount, 37200 Tours, France}\\
\vspace*{5pt}
{\em ${}^{b}$Physics Faculty, Tomsk State University, \\Lenin ave. 36, Tomsk 634050, Russia}\\
\vspace*{5pt}
{\em ${}^{c}$ Service de Physique de l'Univers, Champs et Gravitation, \\ Universit\'e de Mons, 20 place du Parc, 7000 Mons, 
Belgium}\\

\end{center}

\vskip 0.02\textheight

\begin{abstract}
SYK models provide an interesting playground for exploring the $AdS_2/CFT_1$ correspondence. We focus on a class of SYK models that exhibit higher-spin symmetry, whose gravity sector is described by a BF theory generalizing Jackiw--Teitelboim gravity to higher spins. We further develop this framework by constructing consistent interactions between higher-spin gauge fields and scalar matter fields. Two concrete realizations are proposed: Model A, arising from a deformation of the universal enveloping algebra of $\mathfrak{sl}(2,\mathbb{R})$, and Model B, a perturbatively local Poisson sigma-model with an infinite-dimensional target space. While both capture higher-spin dynamics on $(A)dS_2$, they differ in their algebraic structures and locality properties, thus offering complementary perspectives on higher-spin holography.
\end{abstract}

\newpage
\tableofcontents
\newpage

\section{Introduction}
\label{sec:}
\pagenumbering{arabic}
\setcounter{page}{2}
\setcounter{footnote}{0}
Conformal field theories (CFTs) play a central role in modern theoretical physics.
Apart from describing second-order phase transitions in the real world, they arise as fixed points of quantum field theories and as holographic duals of quantum gravity theories with a cosmological constant (most often negative, but sometimes positive).

Heuristically, one may think of every CFT as being associated with some (A)dS dual quantum gravity model. However, the gravitational side is frequently in a deeply quantum regime where no reliable computations can be performed. The large-$N$ regime on the CFT side guarantees that the bulk constituents are semi-classical. Yet even in this regime, the bulk theory may still be intractable -- often because of strong non-locality. In string theory language, the ``easy'' regime corresponds to a weakly coupled worldsheet theory, where the string length is much smaller than the AdS radius, while the ``hard'' regime arises when the string length is comparable to or larger than the AdS radius, making the worldsheet sigma-model strongly curved.

From the perspective of the large-$N$ expansion, all models can be broadly divided into three classes: vector, matrix, and tensor models.
Vector and tensor models are comparatively simple because one can efficiently sum the dominant diagrams at leading (and sometimes all) orders in $1/N$. 
These dominant diagrams are bubbles for vector models and melons for tensor models.
By contrast, matrix models are dominated by planar diagrams, which lack a simple closed-form summation. In certain cases, however, matrix models exhibit integrability, offering an alternative handle. A particularly important subclass of tensor models is the Sachdev–Ye–Kitaev (SYK) model, which introduces an additional element of disorder and has been the focus of intense recent study \cite{Sachdev:1992fk,Kitaev,Maldacena:2016hyu,Gross:2017hcz,Gross:2017vhb,Gross:2017aos,Sarosi:2017ykf,Rosenhaus:2018dtp,Trunin:2020vwy,Sachdev:2024gas,Jha:2025rrz}. One way to distinguish these classes is through anomalous dimensions: they respectively scale as $N$ in $\mathcal{N}=4$ super Yang-Mills (SYM) theory and as $1/N$ in vector models, while they are of order $1$ in SYK models. Out-of-time-ordered correlators (OTOCs) and Lyapunov exponents provide additional diagnostics for separating these classes.
It is remarkable that in each of the three cases (vector, matrix, tensor) there is a regime where a higher-spin gravity description emerges. 

Free or large-$N$ vector models possess infinitely many conserved higher-spin currents. In the free case, these currents are exactly conserved, while in the interacting case they acquire anomalous dimensions through the $1/N$ expansion, so their conservation is slightly broken \cite{Maldacena:2012sf}. According to the AdS/CFT correspondence, higher-spin currents are dual to massless higher-spin fields in the bulk, with the stress tensor corresponding to the graviton. Thus, the AdS dual of a vector model should be a higher-spin gravity theory \cite{Klebanov:2002ja,Sezgin:2003pt,Leigh:2003gk,Giombi:2011kc}. The same bulk  higher-spin theory can be dual to both free and critical (interacting) vector models, depending on the choice of boundary conditions. The price of the relative simplicity of vector models is that their higher-spin AdS dual is necessarily highly non-local \cite{Bekaert:2015tva,Maldacena:2015iua,Sleight:2017pcz,Ponomarev:2017qab}. This  $AdS_{d+1}/CFT_d$ duality between higher-spin gravity theories and vector-models can be formulated in various dimensions and extends to all $d$ for the free boundary theory. 

For matrix models, there are actually two higher-spin regimes. Within the duality between $\mathcal{N}=4$ SYM theory and Type IIB string theory on $AdS_5 \times S^5$, the tensionless limit of the string corresponds to weakly coupled (or even free) SYM \cite{HaggiMani:2000ru,Sundborg:2000wp,Sezgin:2001yf,Mikhailov:2002bp,Beisert:2003te,Beisert:2004di}. As a free $CFT_4$, it possesses higher-spin currents whose $AdS_5$ duals are massless higher-spin fields. However, the bulk description in this regime is extremely complicated: it is a higher-spin gravity coupled to infinitely many multiplets associated with long-trace operators. The higher-spin theory itself suffers from the same non-locality issues as in the vector model case.\footnote{At this level, there is no difference since the higher-spin currents are bilinear in the SYM fields.}

Another regime arises \cite{Chang:2012kt} in the context of ABJ(M) theory \cite{Aharony:2008ug,Aharony:2008gk}. The ABJ theory, which is dual to M-theory on $AdS_4\times S^7/\mathbb{Z}_k$, involves two gauge groups $U(M)_k\times U(N+M)_{-k}$, with matter in the bi-fundamental representation. For large $N$, one can also take $M$ be finite, so that the matter sector effectively reduces to a collection of vector multiplets. In this regime, the theory features higher-spin currents and is expected to be dual to an $\mathcal{N}=6$ $U(M)$-gauged higher-spin gravity theory on $AdS_4$.
A key advantage of this duality is that the higher-spin gravity admits a consistent truncation to chiral higher-spin gravity \cite{Metsaev:1991mt,Metsaev:1991nb,Ponomarev:2016lrm}, whose $AdS_4$ extension has been worked out in \cite{Sharapov:2022awp,Sharapov:2022wpz,Sharapov:2022nps}. This truncated theory can be understood as its self-dual subsector -- analogously to the relation between full Yang–Mills and gravity theories on the one hand, and their self-dual counterparts on the other hand.\footnote{There are, however, important differences: ordinary gravity is non-renormalizable, whereas higher-spin gravity is expected to be UV-finite (once properly defined).} Unlike the full higher-spin gravity theory, chiral higher-spin gravity is perturbatively local, hence, well-defined in the bulk.
Another notable feature is that this $AdS_4/CFT_3$ duality involves interacting theories on both sides, in contrast to the $AdS_5/CFT_4$ example, where an extremely complicated theory in the bulk corresponds to the free SYM theory on the boundary.

For tensor models, and in particular for a large class of SYK-type models, one can identify several theories that exhibit higher-spin symmetry, such as the free or double-scaled SYK models (see e.g. Table \ref{table2} below). In each case, the natural bulk dual to consider is a higher-spin extension of Jackiw--Teitelboim (JT) gravity, as constructed in \cite{Alkalaev:2019xuv,Alkalaev:2020kut}. The \textit{gravity sector} takes the form of a BF theory in which $\mathfrak{sl}(2,\mathbb{R})$ is replaced by its infinite-dimensional higher-spin extension $\mathfrak{gl}[\lambda]$, as in \cite{Alkalaev:2014qpa}. This higher-spin algebra constitutes a one-parameter family of associative algebras, obtained by quotienting the universal enveloping algebra $\mathcal{U}\big(\mathfrak{sl}(2,\mathbb{R})\big)$ of the Lie algebra $\mathfrak{sl}(2,\mathbb{R})$ by the ideal generated by fixing the value of the Casimir element \cite{Feigin}:
\begin{align}\label{undefalg}
        &[h,e_{\pm}]=\pm e_{\pm}\,,\qquad [e_+,e_-]=h\,,
        &e_+e_-+e_-e_+ +h^2-\frac14(\lambda^2 -1)=0\,.
\end{align}
The \textit{matter sector} is built from infinitely many bulk scalar fields, which are dual to an infinite tower of ``single-trace'' operators in the SYK-type model, with scaling dimensions $\Delta = n+1-\lambda$ ($n=0,1,2,\ldots$). At free level, this infinite multiplet naturally furnishes a representation of the higher-spin algebra. The central problem we address in this paper is how to couple the higher-spin gauge fields to this matter sector with  backreaction of the latter on the former.

Since the structure of the interactions in the dual of SYK-like models is not yet clear, we introduce two candidate bulk higher-spin theories, which we call models A and B. These theories have slightly different symmetry algebras; they are nearly indistinguishable at the level of free fields but their physical status differ substantially:

On the one hand, Model A can be viewed as a deformation of the higher-spin algebra extended by a reflection operator $k$ as follows:
\begin{align}\label{MatlambdaZ2deformed1} &[h,e_{\pm}]=\pm e_{\pm}\,,\qquad [e_+,e_-]=h+\nu \,kh\,,\\ &e_+e_-+e_-e_+ +h^2-\frac14(\lambda^2 -1)-\frac{\nu}{2}\,k+\frac{\nu^2}{4}=0\,,\\ &k\,e_{\pm}=-e_{\pm}\,k\,,\qquad k\,h=h\,k\,,\qquad k^2=1\,. \label{MatlambdaZ2deformed3}\end{align}
The first line defines the deformed commutators of $\mathfrak{sl}(2,\mathbb{R})$ for $\nu\neq0$; the second line fixes the value of the quadratic Casimir element; and the third line specifies $k$ as a reflection operator, which plays an essential role in coupling to matter \cite{Alkalaev:2020kut}. At $\nu=0$ and with $k$ omitted, one recovers the undeformed higher-spin algebra \eqref{undefalg}. As we explain in the text, the $\nu$-deformation can be converted into formally consistent non-linear equations of motion, which are integrable. However, unless special tunings are imposed, these equations are likely to exhibit the kind of non-locality characteristic of the higher-spin duals of vector models. Therefore, one can think of the algebra above as an algebraic structure which captures the symmetry of the interactions that is stable under non-local field-redefinitions. A possible action that captures the dynamics (more precisely, the Lax pair) of Model A is the BF-theory for the algebra \eqref{MatlambdaZ2deformed1}, which is a particular example of Poisson sigma-model.

On the other hand, Model B is a perturbatively local field theory formulated as a Poisson sigma-model. That is, it admits an action whose equations of motion take the form
\begin{align}\label{PSMintro}
    d\omegab_k=\tfrac12\partial_k\pi^{ij}(\Cb)\,\omegab_i\,\omegab_j\,, \qquad  \qquad  d\Cb^i=\pi^{ij}(\Cb)\,\omegab_j\,.
\end{align}
Here, $\pi^{ij}(\Cb)$ denotes a Poisson bivector.
A crucial distinction from standard Poisson sigma-models is that the Poisson manifold in this case is infinite-dimensional. This feature allows the model to describe propagating degrees of freedom, preventing it from being purely topological. The Poisson bivector starts with a linear term, i.e. $\pi^{ij}(\Cb)=\pi^{ij}_k \Cb^k+\ldots$\,, which defines a Lie algebra with structure constants $\pi^{ij}_k$. In this instance, the Lie bracket is merely obtained from the Moyal--Weyl commutator of functions on two-dimensional phase space. 
Interestingly, the above construction arises directly from the four-dimensional  chiral higher-spin gravity \cite{Sharapov:2022wpz,Sharapov:2022nps}. Because Model B admits an action and is perturbatively local, it provides a natural framework for computing physical observables in the SYK-type hologram.

The original SYK model, introduced in \cite{Sachdev:1992fk,Kitaev} (see e.g. the
reviews \cite{Sarosi:2017ykf,Rosenhaus:2018dtp,Trunin:2020vwy,Sachdev:2024gas,Jha:2025rrz}), remains a tantalising goal for a complete $AdS_2/CFT_1$ holography since the seminal works \cite{Kitaev,Maldacena:2016hyu}. In particular, elucidating the nature of the bulk dual remains particularly challenging if one aims to reproduce all correlation functions.
The perturbative reconstruction of the bulk duals from the correlation functions of flavor singlets in SYK models was initiated by Gross and Rosenhaus \cite{Gross:2017hcz,Gross:2017vhb,Gross:2017aos} but a simple bulk theory, built from an independent first principle, is still lacking. Nevertheless, several interesting proposals have been made (see e.g. \cite{Jevicki:2016bwu,Jevicki:2016ito,Gross:2017aos,Das:2017hrt,Gaikwad:2018dfc}).
By construction, our models A and B proposed here, must be dual to  SYK-type models which are integrable and whose symmetries are deformations and extensions of $\mathcal{U}\big(\mathfrak{sl}(2,\mathbb{R})\big)$. Note that the correlation functions of the double-scaled SYK model \cite{Berkooz:2018jqr,Lin:2023trc} are known to exhibit a quantum
group symmetry $\mathcal{U}_q\big(\mathfrak{sl}(2,\mathbb{R})\big)$.
This provides some hint that the higher-spin theories proposed here may provide a fruitful source of inspiration for the bulk dual of more challenging SYK models, with broken or $q$-deformed higher-spin symmetries.  

The remainder of the paper is organized as follows. In Section~\ref{sec:2}, we review key aspects of SYK models and identify the features most relevant for their higher-spin gravity duals. In Section~\ref{sec:sl2}, we recall the definitions and structures underlying higher-spin symmetries. Section~\ref{sec:free} introduces the BF-model that captures the dual higher-spin gravity at free level. Our two candidate models, A and B, are then presented and analyzed in Section~\ref{sec:FHSG}. Conclusions and a broader discussion are given in Section~\ref{sec:discussion}. Two technical appendices provide additional details.

\section{SYK models}
\label{sec:2}

\subsection{Field theory}\label{sec:CSYK}

Quite generally, an ``SYK model'' is a theory with a $q$-body interaction of a large number ($N\gg 1$) of Majorana \cite{Sachdev:1992fk,Kitaev} (or Dirac \cite{Sachdev:2015efa,Gu:2019jub}) fermions $\vec\chi(\tau)$ in one ``spacetime'' dimension taking values in the fundamental representation $\mathbb{R}^N$ (or $\mathbb{C}^N$) of the orthogonal group $O(N)$ (or unitary group $U(N)$), with a Gaussian averaging over the coupling constants of some interaction of even degree $q$, which is such that the resulting theory is invariant under the flavor group $O(N)$ (or $U(N)$). 

There are several options for the kinetic term. The most obvious choice is to consider a local kinetic term. Since $\vec\chi$ is Grassmann odd, the only translation and $O(N)$ invariant possibilities are\footnote{In contrast with the SYK literature, we will work in real time (rather than Euclidean time) in order to match higher-spin literature conventions. Note that this may bring imaginary factors and reality conditions with respect to SYK literature, which of course do not affect our mostly kinematical considerations.} 
\begin{align}\label{ON}
S^{\text{loc}}_2[\vec\chi\,;\ell]=  \frac1{(2\ell-1)!}  \int d\tau \, \vec{\chi}(\tau) \cdot (i\pl_\tau)^{2\ell-1} \vec{\chi}(\tau)\,,
\end{align}
where $\ell=1,2,3,\ldots$ The scaling dimension of the fields at the Gaussian fixed point is $\Delta=1-\ell$. The case $\ell=1$ corresponds to the usual SYK model with first-order kinetic term.
More generally, for Dirac fermions the $U(N)$-invariant possibilities are 
\begin{align}\label{UN}
S_2[\vec\chi\,;n]=  \frac1{n!}\int d\tau \, \vec{\chi}^{\,\dagger}(\tau) \cdot (i\pl_\tau)^n\vec{\chi}(\tau)\,,
\end{align}
where $n\in\mathbb{N}$. Therefore, in such cases the scaling dimension of the fields at the Gaussian fixed point is $\Delta=\frac{1-n}2$. 

Gross and Rosenhaus considered \cite{Gross:2017vhb} another generalisation of \eqref{ON} (which we have trivially extended to the case of Dirac fermions) that can admit a line of fixed points, and whose bilocal kinetic term corresponds to the one of a generalised free CFT:
\begin{align}\label{kinetic}
S^{\text{biloc}}_2[\vec\chi\,;\Delta]  &\,=\, i\int_{\tau_1>\tau_2} d\tau_1 \, d\tau_2\, \frac{\vec{\chi}^{\,\dagger}(\tau_1)\cdot \vec{\chi}(\tau_2)}{|\tau_1 -\tau_2|^{2(1-\Delta)}}
\,=\,\frac{i}2\int d\tau_1 \, d\tau_2\, \frac{\text{sgn}(\tau_1-\tau_2)}{|\tau_1 -\tau_2|^{2(1-\Delta)}}\,\vec{\chi}^{\,\dagger}(\tau_1)\cdot \vec{\chi}(\tau_2) 
\end{align}
with $\Delta\in\mathbb R$ the scaling dimension of the fermions $\vec \chi$ at the Gaussian fixed point. 
Note that the kinetic terms \eqref{ON} and \eqref{kinetic} are intimately related, up to a divergent factor, in the sense that\footnote{This follows from the distributional identity (see e.g. Section I.3.3 of \cite{Gelfand}) on their integral kernels
\begin{equation}
    \lim\limits_{\lambda\,\to\, 2\ell-1}\left[\frac{|\tau|^{-(\lambda+1)}\,\text{sgn}(\tau)}{2(\lambda-2\ell+1)}\right]=\frac{\partial_\tau^{2\ell-1}\delta(\tau)}{(2\ell-1)!}
\end{equation}
}
\begin{equation}
    \lim\limits_{\lambda\,\to\, 2\ell-1}\Big((\lambda-2\ell+1)^{-1}S^{\text{biloc}}_2[\vec\chi;\tfrac{1-\lambda}2]\,\Big)=S^{\text{loc}}_2[\vec\chi;\ell]\,.
\end{equation}
We will often use the convenient bra-ket notation with respect to the sesquilinear form
\begin{align}
   \langle \chi_1 |\chi_2\rangle=
\int d\tau_1 \, d\tau_2\, \vec{\chi}_1^{\,\dagger}(\tau_1)\cdot \vec{\chi}_2(\tau_2)
\end{align}
obeying the symmetry property $\overline{\langle \chi_1 |\chi_2\rangle}={\langle \chi_2 |\chi_1\rangle}$.
Let us further introduce the real parameter $\lambda\in\mathbb R$ and set
\begin{equation}
\Delta_-=\Delta\,,\qquad \Delta_+=1-\Delta\,,\qquad \Delta_\pm = \frac{1\pm \lambda}2\,.    
\end{equation}
In this way, both kinetic terms \eqref{UN} and \eqref{kinetic} can be summarised in the same compact form as follows
\begin{align}
S_2[\vec\chi\,;\lambda]&=    \langle \chi | \hat K_\lambda| \chi\rangle=
\int
d\tau_1 \, d\tau_2\, K_\lambda(\tau_1,\tau_2)\,\vec{\chi}^{\,\dagger}(\tau_1)\cdot \vec{\chi}(\tau_2)\,,
\end{align}
where 
\begin{equation}
    K_\lambda(\tau_1,\tau_2)=K_\lambda(\tau_1-\tau_2)=\langle\tau_1|\hat{K}_\lambda|\tau_2\rangle
\end{equation}
is the integral kernel of the kinetic operator $\hat K_\lambda$. The latter operator is Hermitian, i.e. 
\begin{equation}
\hat K_\lambda^\dagger=\hat K_\lambda\;\Longleftrightarrow\; \overline{K_\lambda(\tau_1,\tau_2)}=K_\lambda(\tau_2,\tau_1)\,.    
\end{equation}
More explicitly,
\begin{align}
  K_\lambda(\tau)&=\begin{cases}
        \frac{i}2|\tau|^{-(\lambda+1)}\,\text{sgn}(\tau)\,, & \lambda\notin \mathbb{N}\,,\\
        \frac1{n!}(i\partial_\tau)^n\delta(\tau) \,,& \lambda=n\in \mathbb{N}\,.
    \end{cases}
\end{align}
Let $q\in 2\mathbb N$ be a non-negative even integer. The $q$-body interaction term takes the form
\begin{align}\label{intloc}
    S^{\text{loc}}_q[\vec \chi]&= -\,i^q\int d\tau\, J_{i_1...i_{\frac{q}2}j_1...j_{\frac{q}2}}\, \chi^{\dagger i_1}(\tau)\cdots\chi^{\dagger i_{\frac{q}2}}(\tau)\,\chi^{j_1}(\tau)\cdots\chi^{j_{\frac{q}2}}(\tau)\,,
\end{align}
where the indices $i,j$ take $N$ values and $J_{i_1...i_q}$ is a totally antisymmetric tensor.
Accordingly, at the interacting fixed point the scaling dimension of the fields $\vec\chi$ is $\Delta^{\text{int}}=1/q$. Note that for $q=2$, the two-body interaction is clearly a quadratic term and the model is sometimes called ``nearly-free'', in the sense that it is free before averaging and it keeps many features of the free model.

The latter tensors $J_{i_1...i_q}$ form a collection of coupling constants taken to be independent Gaussian random variables, with zero mean and a variance normalised such that, after averaging, the interaction becomes effectively the bilocal flavor-invariant  term
\begin{align}\label{intnonloc}
    S^{\text{biloc}}_{2q}[\vec \chi\,;J]&= \frac{J^2}{q^2\,N^{q-1}}\int d\tau_1\,d\tau_2\, \left(\vec\chi^{\,\dagger}(\tau_1)\cdot\vec\chi(\tau_2)\right)^q\,,
\end{align}
where $J$ is a dimensionless coupling constant (i.e., at the interacting fixed point, its scaling dimension vanishes).
The case $q=4$ corresponds to the usual SYK model.
The total action is 
\begin{equation}\label{totalact}
S^\text{tot}[\vec \chi\,;J,\lambda,q]=S_2[\vec \chi\,;\lambda]+S^{\text{biloc}}_{2q}[\vec \chi\,;J]\,.
\end{equation}

Let us introduce a bilocal source $H(\tau_1,\tau_2)$ subject to the hermiticity property $\overline{H(\tau_1,\tau_2)}=H(\tau_2,\tau_1)$.
The generating functional $W[H\,;J,\lambda,q]$ of connected correlators of $U(N)$-invariant bilinears $\vec\chi^{\,\dagger}(\tau_1)\cdot\vec\chi(\tau_2)$ will be obtained by performing the path integral over the fundamental field for the action functional 
\begin{equation}\label{2fields}
S[\vec \chi,H\,;J,\lambda,q]=S^{\text{tot}}[\vec \chi\,;J,\lambda,q]+S^{\text{min}}[\vec \chi;H]
\end{equation}
with minimal coupling to the bilocal source $H(\tau_1,\tau_2)$ via
\begin{equation}
S^{\text{min}}[\vec \chi,H]=\int d\tau_1 \, d\tau_2\,H(\tau_1,\tau_2)\,\vec\chi^{\,\dagger}(\tau_1)\cdot\vec\chi(\tau_2)\,.
\end{equation}

In order to perform a Gaussian path integral over the fundamental fields $\vec\chi$, the standard Hubbard-Stratonovich trick is to transform the multilinear interaction term into a bilinear term by introducing two auxiliary (Hermitian, bilocal) fields: $\Sigma(\tau_1,\tau_2)$ and $G(\tau_1,\tau_2)$. The total action becomes:
\begin{eqnarray}
&&S[\vec \chi,H,G,\Sigma\,;J,\lambda,q]=S_2[\vec \chi\,;\lambda]+S^{\text{min}}[\vec \chi,H]\nonumber\\
&&\qquad\qquad -\int d\tau_1 \, d\tau_2\,\Big[\,\Sigma(\tau_1,\tau_2)\Big(\vec\chi^{\,\dagger}(\tau_1)\cdot\vec\chi(\tau_2)-N\,G(\tau_1,\tau_2)\Big)+\frac{J^2N}{q^2}\,G(\tau_1,\tau_2)^q\,\Big]\,.
\end{eqnarray}
By introducing the operators $\hat H$, $\hat\Sigma$ and $\hat G$ whose integral kernels are respectively the bilocal fields $H(\tau_1,\tau_2)$, $\Sigma(\tau_1,\tau_2)$ and $G(\tau_1,\tau_2)$, as well as the operator $\hat{\mathcal{O}}_q$ whose integral kernel is $\langle\tau_1|\hat{\mathcal{O}}_q|\tau_2\rangle=G(\tau_1,\tau_2)^{\frac{q}2}$, the action can be written in a suggestive operatorial way:
\begin{equation}\label{4fields}
S[\vec \chi,H,G,\Sigma\,;J,\lambda,q]=\langle \chi |\,\hat K_\lambda+\hat{H}-\hat\Sigma\,| \chi\rangle+
N\,\text{Tr}\Big[\,\hat\Sigma\,\hat G\,+\tfrac{J^2}{q^2}\,\hat{\mathcal{O}}^2_q\,\Big]\,,
\end{equation}
where
\begin{equation}
    \text{Tr}\Big[\,\hat A\,\hat B\,\Big]=\int d\tau_1 \, d\tau_2\,A(\tau_1,\tau_2)\,B(\tau_2,\tau_1)\,.
\end{equation}
Note that, by further introducing the pure-state density matrix operator $\hat\rho=
  | \chi\rangle \langle \chi |$\,, one can even write all terms in the action \eqref{4fields} in a similar form
\begin{equation}\label{3fields}
S[\vec \chi,G,\Sigma\,;\lambda,q]=
\text{Tr}\Big[\,\hat\rho\,(\hat K_\lambda+\hat{H}-\hat\Sigma)+N\,\Big(\hat\Sigma\,\hat G\,+\tfrac{J^2}{q^2}\hat{\mathcal{O}}^2_q\Big)\Big]\,,
\end{equation}

Performing the Gaussian Berezin integral over the fundamental fields $\vec\chi$ in \eqref{4fields}, leads to the following expression of the effective functional in terms of the bilocal fields
\begin{equation}\label{collectiveFT}
    I[H,G,\Sigma\,;J,\lambda,q]=N\,\text{Tr}\Big[-\,\log(\hat K_\lambda+\hat H-\hat\Sigma)\,+\,\hat\Sigma\,\hat G\,+\,\tfrac{J^2}{q^2}\,\hat{\mathcal{O}}^2_q\,\Big]\,,
\end{equation}
where $1/N$ plays the role of melonic expansion parameter.
This action \eqref{collectiveFT} will be called the \textit{collective field theory} since it only involves the Hubbard-Stratonovich fields $G$ and $\Sigma$ as dynamical fields.
In the large $N$ limit, the generating functional $W[H;J,\lambda,q]$  of connected correlators is obtained by evaluating $I[H,G,\Sigma\,;J,\lambda,q]$ on the solutions of the Schwinger-Dyson equations
\begin{equation}\label{eomcollective}
    \Sigma(\tau_1,\tau_2)=-\frac{J^2}{q^2}\,G(\tau_1,\tau_2)^{q-1}\,,\qquad \hat{G}=(\hat K_\lambda+\hat H-\hat\Sigma)^{-1}\,,
\end{equation}
so that $\Sigma$ takes the interpretation of a self-energy and $G$ of a dressed propagator.
For holographic duality to hold at the level of correlation functions, the large-$N$ limit of the generating functional $W[H;J,\lambda,q]$  should coincide with the on-shell action of some bulk gravity functional with  bilocal fields $H$ as boundary data.

\subsection{Symmetries}\label{sec:symSYK}

By construction, the total action \eqref{totalact} is invariant under the flavor group $O(N)$ or $U(N)$. Aside from this internal symmetry, the above kinetic terms \eqref{ON}-\eqref{kinetic} all have conformal symmetry $\mathfrak{so}(2,1)$ whose
 generators $\hat P$, $\hat C$ and $\hat D$  -- representing translation, conformal boost and dilatation, respectively -- are given by the first-order differential operators:
\begin{align}\label{conformalgen}
    \hat P&= \pl_t\,, & \hat C&=t^2 \pl_t +2\Delta\, t\,, & \hat D&=t\,\pl_t+\Delta  \,,
\end{align}
where the corresponding scaling dimension $\Delta=\Delta^{\text{free}}=\tfrac{1-\lambda}2$ was mentioned above. Similarly, the interaction terms \eqref{intloc} and \eqref{intnonloc} are invariant under the algebre $\mathfrak{so}(2,1)$ spanned by generators \eqref{conformalgen} with $\Delta=\Delta^{\text{int}}=1/q$. Therefore, one expects a line of  fixed points  when $\Delta^{\text{free}}=\Delta^{\text{int}}$, cf. the model \cite{Gross:2017vhb} dubbed ``cSYK'', which corresponds to the case $\lambda=1-\frac2{q}$.
See Table \ref{table1} for a summary of the relevant SYK fixed points.

\begin{table}[h!]
\begin{center}
	\begin{tabular}{|c|c|c|}
    \hline
	Fixed point & Scaling dimension  & Parameter $\lambda$ \\\hline\hline
    Gaussian (bilocal) & $\Delta^{\text{free}}=\frac{1-\lambda}2$ & $\lambda\in\mathbb R$
	\\\hline
    Gaussian (local) & $\Delta^{\text{free}}=\frac{1-n}2$ & $n\in\mathbb N$
	\\\hline
    Gaussian (SYK) & $\Delta^{\text{free}}=0$ & $1$
	\\\hline\hline
    Interacting ($q$-body) & $\Delta^{\text{int}}=\frac1{q}\in [0,\frac12]$ & $\lambda=1-\frac{2}{q}\in [0,1]$
	\\\hline
    Interacting (2-body) & $\Delta^{\text{int}}=\frac12$ & $0$
	\\\hline
    Interacting (SYK) & $\Delta^{\text{int}}=\frac14$ & $\frac12$
	\\\hline
Interacting ($q=\infty$) & $\Delta^{\text{int}}=0$ & $1$
	\\\hline\hline
Line (cSYK) & $\Delta^{\text{free}}=\Delta^{\text{int}}=\frac1{q}$ & $\lambda=1-\frac{2}{q}$
	\\\hline
	\end{tabular}
\end{center}
\caption{\label{table1}Scaling dimensions $\Delta$ for various relevant SYK fixed points.}
\end{table}

By usual arguments in higher-spin literature (see e.g. \cite{Mikhailov:2002bp,Bekaert:2006zoe}), one can show that the symmetry under the conformal algebra $\mathfrak{so}(2,1)$ is enhanced, for quadratic terms, to its enveloping algebra (which will be discussed at length in Section \ref{sec:sl2}). In other words, the quadratic terms, such as the kinetic term (or the ``interaction'' term \eqref{intloc} when $q=2$), are invariant under all infinitesimal higher-spin symmetries generated by higher-order differential operators that are polynomials of the generators \eqref{conformalgen}.

Let us anticipate the group-theoretical discussion in Section \ref{sec:sl2} to recall some basic facts: according to \eqref{conformalgen} the field $\vec\chi$ with scaling dimension $\Delta=\Delta^{\text{free}}=\tfrac{1-\lambda}2$ is a primary field which, together with the tower of all its derivatives, spans the Verma module $\mathcal{V}_{\tfrac{1-\lambda}2}$ of the Lie algebra $\mathfrak{so}(2,1)$.
For $\Delta>0\Leftrightarrow \lambda<1$, this is a unitary irreducible representation of 
$\mathfrak{so}(2,1)$
belonging to the discrete series (see e.g. \cite{Kitaev:2017hnr} for a review). This is for instance the case at the interacting fixed point since $\lambda=1-\frac{2}{q}$ and $q>0$. If $\lambda=-n$ with $n\in\mathbb N$, then the unitary irreducible representation $\mathcal{V}_{\tfrac{1+n}2}$ of the Lie algebra
$\mathfrak{so}(2,1)\simeq\mathfrak{sl}(2,\mathbb{R})$ lifts to a unitary irreducible representation of the Lie group $SL(2,\mathbb{R})$.
Note that, for the local kinetic term \eqref{UN}, the primary field $\vec\chi$ of scaling dimension $\Delta_-=\frac{1-n}2$ and its descendants span, off-shell, the Verma module $\mathcal{V}_{\frac{1-n}2}$. On-shell, the primary field obeys to the equation of motion $\partial_\tau^n\vec\chi=0$. The left-hand-side is the descendant $\partial_\tau^n\vec\chi$ of the primary field $\vec\chi$ of scaling dimension $\Delta_-=\frac{1-n}2$ which is itself also a primary field $\partial_\tau^n\vec\chi$, of scaling dimension $\Delta_+=\frac{1+n}2$ and which defines a Verma submodule $\mathcal{V}_{\frac{1+n}2}\subset\mathcal{V}_{\frac{1-n}2}$. On-shell, the fields span the finite-dimensional module $\mathcal{D}_{\frac{n-1}2}\simeq\mathcal{V}_{\frac{1-n}2}/\mathcal{V}_{\frac{1+n}2}$ of dimension $n$. Accordingly, the on-shell higher-spin symmetry algebras are finite-dimensional in these cases (see Section \ref{sec:sl2}).

Consider now the Weyl transformations, i.e. $d\tau\to d\tau'=\Omega(\tau)\,d\tau$ with $\Omega$ an everywhere positive function. The interaction terms \eqref{intloc} and \eqref{intnonloc} are invariant under 
these Weyl transformations, provided the fields $\vec\chi$ transform as conformal densities of suitable weight, i.e. $\vec\chi'(\tau)=\Omega(\tau)^{-\frac1{q}}\,\vec\chi(\tau)$.
Consequently, the interaction terms \eqref{intloc} and \eqref{intnonloc} are invariant under 
the time reparametrisations $\tau\to\tau'=f(\tau)$ provided the fields $\vec\chi$ transform as usual densities with suitable weight, i.e. $\vec\chi'(\tau')=\big|\frac{\,d\tau'}{\,d\tau\,}\big|^{-\frac1{q}}\,\vec\chi(\tau)$.
The Lie algebra of such infinitesimal diffeomorphisms is generated by first order differential operators of the form $f(\tau)\partial_\tau+\frac1{q}\frac{df(\tau)}{d\tau}$.

For $\ell=1$ or $n=1$, the quadratic actions \eqref{ON} or \eqref{UN} are also invariant under Weyl transformations and time reparametrisations, but with the condition that fields $\vec\chi$ transform as scalar fields (i.e. densities of weight zero). The Lie algebra of the corresponding infinitesimal diffeomorphisms is thus merely generated by vector fields $f(\tau)\partial_\tau$.
Again considering the enveloping algebra of the previous infinitesimal generators lead to the Lie algebra of differential operators. They extend the global higher-spin symmetries (to be discussed at length in Section \ref{sec:sl2}) to local higher-spin symmetries for individual quadratic terms in these partical cases.

By-now usual arguments in higher-spin literature (see e.g. \cite{Segal:2002gd,Bekaert:2010ky}) allow to extend global higher-spin symmetries to local higher-spin symmetries for quadratic actions minimally-coupled to a bilocal source, when the latter is allowed to transform suitably. We will specifically consider the case $q=2$ which undergoes higher-spin symmetry enhancement (as mentioned in \cite{Alkalaev:2019xuv} and made explicit in \cite{Anninos:2023lin}) since the free case can be obtained by simply setting $J=0$ in the formulae below. It is remarkable that the total action \eqref{2fields}, which is the sum of a quadratic and a quartic piece, is invariant under local higher-spin symmetries (see  \cite{Anninos:2023lin} for a similar investigation of these local higher-spin symmetries) for suitable transformation of the background field. A compact way to see this fact is to notice that, for $q=2$, the functional \eqref{2fields} can be written in compact form as
\begin{equation}
S[\vec \chi,H\,;J,\lambda,q=2]\,=\,\delta_{ij}\,\langle \chi^i |\,\hat K_\lambda+\hat{H}\,| \chi^j\rangle\,-\,\frac{J^2}{4N}\,\delta_{ij}\,\delta_{kl}\,\langle \chi^i|\chi^k\rangle \,\langle \chi^l|\chi^j\rangle\,=\,\text{Tr}\Big[\,\hat\rho\,(\hat K_\lambda+\hat{H})-\frac{J^2}{4N}\,\hat{\rho}^2\Big]\,,
\end{equation}
This functional is manifestly invariant under the following higher-spin symmetries
\begin{equation}\label{HSU}
    |\chi^i\rangle\to \hat U|\chi^i\rangle\,,\qquad \hat\rho\to \hat U\hat \rho\,\hat U^{-1}\,,\qquad \hat K_\lambda+\hat{H}\to \hat U(\hat K_\lambda+\hat{H})\hat U^{-1}\,,
\end{equation}
where $\hat U$ is a unitary operator (with respect to the $L^2$ norm on the time domain): $\hat U^\dagger=\hat U^{-1}$.
For $q=2$, one can also explicitly compute the collective field theory \eqref{collectiveFT}:
\begin{equation}
    I[H,G,\Sigma\,;J,\lambda,q=2]=N\,\text{Tr}\Big[-\,\log(\hat K_\lambda+\hat H-\hat\Sigma)\,+\,\hat\Sigma\,\hat G\,-\,\frac{J^2}{4}\,\hat G^2\,\Big]\,,
\end{equation}
since $\hat{\mathcal{O}}_2=\hat G$. The auxiliary field $G(\tau_1,\tau_2)$ can easily be eliminated via the solution $G=\tfrac{2}{J^{2}}\Sigma$ of its own equation of motion. This leads to
\begin{equation}\label{eomI}
    I[H,\Sigma\,;J,\lambda,q=2]=N\,\text{Tr}\Big[-\,\log(\hat K_\lambda+\hat H-\hat\Sigma)\,+\,\,\tfrac1{J^2}\,\hat \Sigma^2\,\Big]\,,
\end{equation}
which is manifestly invariant under
\begin{equation}
    \hat\Sigma\to \hat U\hat \Sigma\,\hat U^{-1}\,,\qquad \hat K_\lambda+\hat{H}\to \hat U(\hat K_\lambda+\hat{H})\hat U^{-1}\,,
\end{equation}
Let us stress that, strictly speaking, the original total action \eqref{totalact}, i.e. in the absence of source, is not invariant under the local higher-spin gauge symmetries \eqref{HSU} since the source is fixed to zero, $\hat{H}=0$. Thus the original action \eqref{totalact} is only invariant under the global higher-spin symmetries with $\hat U$ obeying to 
\begin{equation}
\hat U\hat K_\lambda\hat U^{-1}    =K_\lambda\,,
\end{equation}
which reduces to the higher-spin symmetries mentioned at the beginning of this subsection. The extra term obtained for the variation under  local higher-spin gauge symmetries \eqref{HSU} could be interpreted as a sort of ``higher-spin Schwarzian action'' of the bilocal gauge parameter $U(\tau_1,\tau_2)$ associated to the unitary operator $\hat U$.
More precisely, in the saddle point approximation ($N=\infty$) one can consider the following variation
\begin{equation}
    I^{\text{Sch}}[\,U\,;J,\lambda,q=2]\,=\,\text{Tr}\Big[\log(\hat K_\lambda-\hat\Sigma_*)-\,\log(\hat K_\lambda-\hat U\hat\Sigma_*\hat U^{-1})\,\Big]\,,
\end{equation}
where $\hat\Sigma_*$ is a solution of the classical equation of motion, $J^2(\hat K_\lambda-\hat\Sigma_*)^{-1}=\hat\Sigma_*$, for the collective field theory \eqref{eomI}.

Due to the integrability of the  SYK model at $q=\infty$, one  also expects to have  higher-spin symmetries similar to the ones at $q=2$. In fact, one may argue that the SYK model exhibits higher-spin symmetry enhancement in the double scaling limit, i.e. in the simultaneous limits $N\to\infty$ and $q\to\infty$ with the ratio $e^2=q^2/N$ fixed. In Appendix \ref{app:Liouville}, we review that the double scaling limit of the collective field theory \eqref{3fields} leads to a two-dimensional Liouville theory \cite{Cotler:2016fpe} (with an elusive symmetry condition on the Liouville field).
Since Liouville theory is known to have higher-spin symmetries (see e.g. \cite{Zhiber,Kiselev} for detailed discussions of the huge symmetries of Liouville theory), in this sense, so does the double scaling limit of SYK model. Incidentally, note that the latter is expected to possess a bulk dual in agreement with $dS_2/CFT_1$ correspondence (see e.g. \cite{Lin:2022nss,Narovlansky:2023lfz,
Verlinde:2024zrh,Blommaert:2024ymv} and refs therein).

\subsection{Bilinear singlet sector}\label{singletsector}

The bilocal $U(N)-$singlet $\vec\chi^{\,\dagger}(\tau_1)\cdot\vec\chi(\tau_2)$
can be thought of as the generating function of the tower of local $U(N)$-singlets 
\begin{equation}\label{bilinearsinglets}
\mathcal{O}_n(\tau)=    \vec\chi^{\,\dagger}(\tau)\cdot\Big(\stackrel{\leftrightarrow}{\partial}_\tau\Big)^n\vec\chi(\tau)\,,\qquad n=0,1,2,\ldots
\end{equation}
in the sense that
\begin{equation}
    \vec\chi^{\,\dagger}(\tau_1)\cdot\vec\chi(\tau_2)=\sum\limits_{n=0}^\infty\frac1{n!}\left(\tfrac{\tau_1-\tau_2}2\right)^n\mathcal{O}_n\left(\tfrac{\tau_1+\tau_2}2\right)\,,
\end{equation}
where
$\chi_1\stackrel{\leftrightarrow}{\partial}\chi_2:=({\partial}\chi_1)\,\chi_2-\chi_1\,({\partial}\chi_2)$.
For Majorana fermions $\vec\chi^{\,\dagger}=\vec\chi$, only the $O(N)$-singlets $\mathcal{O}_n(\tau)=    \vec\chi(\tau)\cdot(\stackrel{\leftrightarrow}{\partial}_\tau)^n\vec\chi(\tau)$ with even spin $n$ are non-vanishing. Note that the local bilinears \eqref{bilinearsinglets} have bare scaling dimensions 
\begin{equation}
    \Delta^{\text{free}}_n=2\Delta+n\,.
\end{equation}
In particular, at the Gaussian SYK model with first-order kinetic term the scaling dimensions
$\Delta^{\text{free}}_n\stackrel{\Delta\to 0}{\to}n$
span all non-negative integer values.\footnote{As a side technical remark, let us mention that the local bilinears \eqref{bilinearsinglets} 
are not primary fields, although they have definite scaling dimensions. Nevertheless, out of them and their descendants, one can cook up a collection of bilinear singlets involving the same number of derivatives which are primary fields (see Appendix A of \cite{Gross:2017hcz} for the explicit expression). We ignore this subtlety here since they are in one-to-one correspondence with each other, as follows from group theory (cf. Section \ref{sec:sl2}).}

Note that at the interacting fixed point of the $q=2$ SYK model, the scaling dimension of $\vec\chi$ is $1/2$ and the singlet bilinears \eqref{bilinearsinglets} have scaling dimension $\Delta^{\text{int}}_n\stackrel{q\to 2}{\to}1+n$ \cite{Pethybridge:2024qci}. In fact, the bare dimension is not corrected since the theory is nearly-free. Thus, together with their descendants they span the whole discrete series of unitary irreducible representations of $SO(2,1)$, a feature which is important for $dS_2/CFT_1$ holography (because only integer eigenvalues are allowed since $dS_2\cong \mathbb{R}\times S^1$ while the corresponding circle is timelike and unwrapped for $AdS_2\cong \mathbb{R}\times\mathbb{R}$) as pointed out in \cite{Anninos:2023lin,Pethybridge:2024qci}.
Similarly, note that at the interacting fixed point of the $q=\infty$ SYK model, the scaling dimension of $\vec\chi$ vanishes and the singlet bilinears \eqref{bilinearsinglets} have scaling dimension $\Delta^{\text{int}}_n\stackrel{q\to\infty}{\to}n$ \cite{Gross:2017hcz}. 

To finish this section, let us summarise the list of integrable SYK models with unbroken higher-spin symmetry. There is (1) the free local models \eqref{UN} for any order of the kinetic term, (2) the generalised free models \eqref{kinetic} for any value of the scaling dimension, (3) the nearly-free model at $q=2$ and (4) the double scaling limit of SYK model. They are summarised, together with their singlet spectrum, in Table \ref{table2}.

\begin{table}[h!]
\begin{center}
	\begin{tabular}{|c|c|c|c|}
    \hline
	Integrable & Definition & Scaling dimensions & Parameter \\
    fixed points & (as SYK models) & (singlet bilinears) & $\lambda$ \\\hline\hline
    Generalized free SYK & $J=0$ & $\Delta^{\text{free}}_n=n+1-\lambda$	& $\lambda\in\mathbb R$ \\\hline
    Free SYK & $J=0$ & $\Delta_n^{\text{free}}=n$ & $1$ 
	\\\hline
    Nearly-free SYK & $q=2$ & $\Delta_n^{\text{int}}=n+1$ & $0$  
	\\\hline
    Double scaled SYK & $N=\infty$, $q=\infty$ & $\Delta_n^{\text{int}}=n$ & $1$  	\\\hline
	\end{tabular}
\end{center}
\caption{\label{table2}Various fixed points   with unbroken higher-spin symmetry, together with their scaling dimensions $\Delta_n$ for singlet bilinears $\mathcal{O}_n$ and the corresping value sof the parameter $\lambda$.}
\end{table}

\section{Some facts about $\mathfrak{sl}(2,\mathbb{R})$ and $\mathfrak{hs}[\lambda]$}
\label{sec:sl2}

In this section, we review basic facts about higher-spin symmetries and their representations, which will allow us to construct a simple BF-model describing free two-dimensional higher-spin gravity around AdS$_2$ background. Since the relevant higher-spin algebra is based on $\mathfrak{sl}(2,\mathbb{R}) \simeq \mathfrak{so}(1,2)$, we begin by summarizing the necessary facts.

Higher-spin algebras are typically infinite-dimensional associative algebras, whose commutator Lie algebras contain the spacetime symmetry algebra $\mathfrak{g}$ as a subalgebra. A natural construction of higher-spin algebras is given by the quotient $\mathcal{U}(\mathfrak{g})/\mathcal{I}$ of the universal enveloping algebra $\mathcal{U}(\mathfrak{g})$ of $\mathfrak{g}$ by a two-sided ideal $\mathcal{I}$. Often, the ideal $\mathcal I$ is taken to be the annihilator $\mathrm{Ann}(V)$ of an irreducible $\mathfrak g$-module $V$. Such two-sided ideals are called \textit{primitive}. For instance, if $V$ is an irreducible representation, then the Casimir elements $C_k\in \mathcal{Z}\big(\mathcal{U}(\mathfrak{g})\big)$ -- which generate the center of $\mathcal{U}(\mathfrak{g})$ -- act as scalars on $V$. That is, $C_k$ takes a fixed value $c_k \in \mathbb{C}$, so that $(C_k - c_k)$ annihilates $V$, and thus $\mathcal{U}(\mathfrak{g})(C_k - c_k) \subset \mathrm{Ann}(V)$.\footnote{For a more detailed discussion, see e.g. \cite{dixmier1996enveloping}.}

In our case, $\mathfrak{g}=\mathfrak{sl}(2,\mathbb{R})$, and the above construction simplifies dramatically. In the Cartan--Weyl basis, the commutation relations of $\mathfrak{sl}(2,\mathbb{R})$ take the form
\begin{align}\label{comrelsl2}
    [\,h,e_{\pm}\,]=\pm \,e_{\pm}\,,\qquad [\,e_+,e_-\,]=h\,.
\end{align}
Let us denote by $\mathcal{V}_{\Delta}$ the Verma module of $\mathfrak{sl}(2,\mathbb{R})$ with lowest weight $\Delta$:
\begin{align}
    e_-|\Delta\rangle&=0 \,,& h|\Delta\rangle=\Delta|\Delta\rangle\,.
\end{align}
The Verma module $\mathcal{V}_{\Delta}$ of $\mathfrak{sl}(2,\mathbb{R})$ is a reasonable candidate for the $\mathfrak{g}$-module $V$. In fact, the Verma module  $\mathcal{V}_{\Delta}$ is unitary and irreducible for $\Delta> 0$. The weights of (non-unitary) finite-dimensional representations are $\Delta=\tfrac12(1-n)$, for integer $ n=1,2,3,\ldots$ In the latter case, $\mathcal{V}_{\Delta}$ is reducible and the quotient $\mathcal{V}_{\Delta}/\mathcal{V}_{1-\Delta}$ is a finite-dimensional irreducible representation $\mathcal{D}_j$ of dimension $ n$, indexed by $j=\tfrac12( n-1)$. In field-theoretic terms, the vector space $\mathcal{D}_{\tfrac{ n-1}{2}}$ is spanned by solutions to the field equations $\pl^ n_\tau \chi_\Delta(\tau)=0$, cf. the paragraph in Section \ref{sec:symSYK}. 

The tensor square of the Verma module $\mathcal{V}_{\Delta}$ decomposes as follows:
\begin{align}\label{tensorproduct}
    \mathcal{V}_{\Delta}\otimes \mathcal{V}_{\Delta}&= \bigoplus_{n=0}^{\infty} \mathcal{V}_{2\Delta+n}\,.
\end{align}
In particular, for $\Delta=\tfrac12(1-\lambda)$ with $\lambda\in\mathbb{R}$ the right-hand side gives the spectrum of primary fields with $\Delta_n= n+1-\lambda$. For finite-dimensional representations, we have the Clebsch--Gordan decomposition
\begin{align}\label{tensorproductfinite}
    \mathcal{D}_{\tfrac{ n-1}{2}}\otimes \mathcal{D}_{\tfrac{ n-1}{2}}&= \bigoplus_{m=0}^{ n-1} \mathcal{D}_{m}\,.
\end{align}
Following \cite{Feigin}, we can now define a one-parameter family of associative algebras,\footnote{We will follow the notations in \cite{Alkalaev:2020kut}.} 
\begin{align}\label{Feiginalg}
    \mathrm{Mat}[\lambda]&= \mathcal{U}\big(\mathfrak{sl}(2,\mathbb{R})\big)/\mathcal{I}_\lambda \,,
\end{align}
where $\mathcal{I}_\lambda=\mathrm{Ann}(\mathcal{V}_\Delta)$ with $\Delta=\Delta_\pm=\tfrac12(1\pm\lambda)$. Notably, the ideal $\mathcal{I}_\lambda$ is generated by a single element $C_2-\tfrac14 (\lambda^2-1)$, where 
\begin{align}\label{C_2}
    C_2\big(\mathfrak{sl}(2,\mathbb{R})\big)= e_+e_-+e_-e_++h^2\in \mathcal{Z}\big(\mathfrak{sl}(2,\mathbb{R})\big)
\end{align} is the quadratic Casimir element of the Lie algebra $\mathfrak{sl}(2,\mathbb{R})$. The associative algebra $\mathrm{Mat}[\lambda]$ will be called \textit{higher-spin algebra}.\footnote{We will sometimes refer to its Lie algebra counterpart $\mathfrak{gl}[\lambda]$ also as the ``higher-spin algebra''. This terminology will not be problematic since they are intimately related and, anyway, the notation disambiguates them.} It is an infinite-dimensional associative algebra, which is simple for generic values of $\lambda$. Its commutator Lie algebra $\mathfrak{gl}[\lambda]=\mathrm{Lie}\big(\mathrm{Mat}[\lambda]\big)$ decomposes into the direct sum of its one-dimensional centre $\mathfrak{u}(1)$, associated with the unit of $\mathcal{U}(\mathfrak{sl}(2,\mathbb{R}))$, plus a Lie algebra denoted by $\mathfrak{hs}[\lambda]$. The latter is simple for generic values of $\lambda$. However, for integer values $\lambda= n\in \mathbb{N}$, the Lie algebra $\mathfrak{gl}[\lambda]$ acquires an infinite-dimensional ideal, with the corresponding quotient being the finite-dimensional general linear algebra $\mathfrak{gl}( n,\mathbb{R})$.
Similarly, the quotient of $\mathfrak{hs}[\lambda]$ for $\lambda= n\in \mathbb{N}$ is the special linear algebra $\mathfrak{sl}( n,\mathbb{R})$.

Since $\mathfrak{sl}(2,\mathbb{R})\subset \mathfrak{gl}[\lambda]$, one can ask about the decomposition of $\mathfrak{gl}[\lambda]$ with respect to various $\mathfrak{sl}(2,\mathbb{R})$-actions. These include the left, right,  and adjoint actions. Under the adjoint action, we have
\begin{align}\label{decompgl1}
    \mathfrak{gl}[\lambda] =\bigoplus_{n=0}^{\infty}\mathcal{D}_{n}\qquad \text{and}\qquad\mathfrak{hs}[\lambda] =\bigoplus_{n=1}^{\infty}\mathcal{D}_{n}\,.
\end{align}
For $\lambda\in\mathbb{N}$, the embedding of $\mathfrak{sl}(2,\mathbb{R})$ inside the finite-dimensional quotients corresponds to the principal embedding, and we find
\begin{align}\label{decompgl2}
    \mathfrak{gl}( n,\mathbb{R}) =\bigoplus_{m=0}^{ n-1}\mathcal{D}_{m}\qquad \text{and}\qquad\mathfrak{sl}( n,\mathbb{R}) =\bigoplus_{m=1}^{ n-1}\mathcal{D}_{m}\,.
\end{align}

One realization of $\mathfrak{sl}(2,\mathbb{R})\simeq\mathfrak{so}(2,1)$ is the conformal one \eqref{conformalgen}. 
The value of the quadratic Casimir operator is fixed as follows:
\begin{align}
    \hat C_2\big(\mathfrak{so}(2,1)\big)=\hat D^2-\tfrac12 (\hat P\hat C+\hat C\hat P)=\Delta(\Delta-1) =-\Delta_+\Delta_-=\tfrac14(\lambda^2-1)\,.
\end{align}
Thus, $\mathrm{Mat}[\lambda]$ can be understood as the associative algebra generated by the first-order differential operators $\hat P$, $\hat C$, and $\hat D$. 

An alternative realization of $\mathrm{Mat}[\lambda]$ can be given for $\lambda=1/2$.  Consider the polynomial Weyl algebra $A_1$, defined on two generators $y_1$ and $y_2$ that satisfy the commutation relation $[y_2,y_1]=2$.\footnote{Setting $y_2=\sqrt{2}a$, $y_1=\sqrt{2}a^\dag$, one can recover the usual creation/annihilation operators. This explains why, in the physical 
literature, the Weyl algebra $A_1$ is often referred to as the \textit{oscillator algebra}.} 
It is straightforward to see that the quadratic polynomials
\begin{align}
    t_{AB}&=-\tfrac14 (y_Ay_B +y_By_A)\,,\qquad A,B=1,2\,,
\end{align}
generate the Lie algebra $\mathfrak{sl}(2,\mathbb{R})$ through their commutators:
\begin{align}\label{osccomA}
    [t_{AB},t_{CD}]&= \epsilon_{AC}\,t_{BD}+\epsilon_{BC}\,t_{AD}+\epsilon_{AD}\,t_{BC}+\epsilon_{BD}\,t_{AC}\,.
\end{align}
The Cartan--Weyl generators are given by
$$e_{+}=\tfrac{i}{2\sqrt2} \,t_{11}\,,\qquad h=-\tfrac12\, t_{12}\,, \qquad e_-=\tfrac{i}{2\sqrt2}\, t_{22}\,.$$ 
The quadratic Casimir operator is fixed as $C_2=-3/16$, which corresponds to $\lambda=1/2$. This demonstrates that $\mathrm{Mat}[\frac12]$ is isomorphic to the subalgebra $A^{\text{even}}_1$ of the Weyl algebra $A_1$ spanned by even polynomials in  $y_A$.\footnote{There is yet another realization of $\mathrm{Mat}[\lambda]$ in terms of a deformed oscillator algebra \cite{Wigner1950,Yang:1951pyq, Boulware1963, Mukunda:1980fv, Vasiliev:1989re}, which allows  $\lambda$ to take any value. }

\section{Higher-spin gravity in two dimensions without backreaction}
\label{sec:free}

In this section, we review how higher-spin algebras allow one to construct a simple BF-model describing free two-dimensional higher-spin gravity around (A)dS$_2$ background. For more details, see \cite{Alkalaev:2013fsa,Grumiller:2013swa,Alkalaev:2014qpa,Alkalaev:2020kut}.

\subsection{Jackiw--Teitelboim gravity}
\label{sec:JT}
The higher-spin gravity we are looking for is an extension of Jackiw--Teitelboim (JT) gravity, which can be formulated as a BF-theory. Indeed, as shown in \cite{Fukuyama:1985gg}, Jackiw--Teitelboim gravity can be described by the action
\begin{align}
    S_{JT}[\phi,\omega]&= \int \phi_{AB} \, R^{AB}\,.
\end{align}
Here 
\begin{align}
    R^{AB}=d\omega^{AB}-\omega\fud{A}{C}\wedge \omega^{CB}
\end{align}
is the curvature two-form for an $\mathfrak{sl}(2,\mathbb{R})$ connection one-form $\omega^{AB}$
and $\phi_{AB}$ is a scalar field, i.e., a zero-form, taking values in the coadjoint representation of $\mathfrak{sl}(2,\mathbb{R})$. The latter originates from an $SO(1,2)$-vector $\phi_a$, which has been translated into the $SL(2,\mathbb{R})$-language. The corresponding equations of motion 
\begin{align}
    R^{AB}&=0\,, & D\phi_{AB}&=0
\end{align}
are simply the flatness and covariant constancy conditions.
\subsection{Higher-spin extension of Jackiw--Teitelboim gravity}
\label{sec:HSJT}

To construct a higher-spin extension of the above theory, the recipe is quite elegant \cite{Alkalaev:2013fsa,Grumiller:2013swa,Alkalaev:2014qpa,Alkalaev:2020kut}: one just replaces $\mathfrak{sl}(2,\mathbb{R})$ with $\mathfrak{sl}( n,\mathbb{R})$, or with the infinite-dimensional algebra $\mathfrak{hs}[\lambda]$ (or even further with an extended algebra $\mathfrak{ehs}[\lambda]$, see below, if one wishes to include propagating matter fields beyond the topological gravitational sector). The off-shell fields of the higher-spin extension of Jackiw-Teitelboim gravity consist of a connection one-form $\omega$ and a zero-form $\phi$ both valued in the considered algebra, say $\mathfrak{sl}( n,\mathbb{R})$ or $\mathfrak{hs}[\lambda]$. 

An additional useful property of the associative unital algebra $\mathrm{Mat}[\lambda]$, inherited from $\mathcal{U}\big(\mathfrak{sl}(2,\mathbb{R})\big)$ and crucial for constructing an action, is that the projection $\Tr:\mathrm{Mat}[\lambda]\to\mathbb{R}$ of the higher-spin algebra onto its unit is a trace, i.e., $\Tr\,[a,b]_\star=0$, where $[a,b]_\star=a\star b-b\star a$ and $\star$ denotes the product in the higher-spin algebra $\mathrm{Mat}[\lambda]$. For $\lambda\in \mathbb{N}$, all these properties are evident for the associative algebra $\text{Mat}_ n$ of square matrices; for example, the trace is the usual matrix trace. 

Remarkably, this trace is a non-degenerate linear form on the higher-spin algebra $\mathrm{Mat}[\lambda]$.
Hence, the action for the higher-spin extension of JT gravity can be written as\footnote{Note that, strictly speaking, a trace is not necessary for writing down a BF-action since it is enough to assume that the zero-form $\phi$ takes values in the coadjoint representation and to make use of the natural pairing between $\mathfrak{g}$ and $\mathfrak{g}^*$. This observation is useful because in this way it is possible to take the In\"onu-Wigner contraction of $\mathfrak{g}$ in any BF-action and obtain a non-trivial action since the pairing remains non-degenerate, while invariant traces often becomes degenerate. In particular, it
is possible \cite{Pannier} to take the flat limit of the higher-spin gravity action in \cite{Alkalaev:2020kut}.}
\begin{align}
    S_{HS-JT}[\phi,\omega]&= \Tr\int \phi \star R\,.
\end{align}
The corresponding equations of motion 
\begin{equation}\label{eoms}
    R=d\omega-\omega\star\omega =0\,,\qquad D\phi=d\phi-\omega\star\phi+\phi\star \omega=0
\end{equation}
can be solved in pure gauge form as $\omega=-g^{-1}\star dg$ and $\phi=-g^{-1}\star \phi_0\star g$ where $\phi_0$ is a constant element of the considered Lie (sub)algebra, e.g. $\mathfrak{sl}( n,\mathbb{R})$  or $\mathfrak{hs}[\lambda]$. This shows that, locally, the solution space is isomorphic to the linear space of the considered algebra. 

Field-theoretically, when linearized about a gravitational background, i.e. locally $(A)dS_2$, the irreducible $\mathfrak{sl}(2,\mathbb{R})$ components of $\omega$ describe partially-massless fields of maximal depth \cite{Alkalaev:2013fsa,Alkalaev:2014qpa}, which is consistent with the dictionary of \cite{Skvortsov:2006at}. Similarly, the irreducible $\mathfrak{sl}(2,\mathbb{R})$ components of $\phi$ correspond to $(A)dS_2$ Killing tensors  that span  the corresponding irreducible $\mathfrak{sl}(2,\mathbb{R})$ modules in \eqref{decompgl1} or \eqref{decompgl2}. 

\subsection{Adding matter}
\label{sec:HSJT+m}

The next step is to add matter to the higher-spin extension of JT gravity. The matter sector  consists of infinitely many scalar fields in accordance with the spectrum of singlets in integrable SYK models with unbroken higher-spin symmetry. A single scalar field on $AdS_2$ obeys
\begin{align}\label{scalarfields}
    (\square +m^2)\varphi=0\,, &&m^2
    =-\Delta_+\Delta_-\,,
\end{align}
which also relates the mass $m$ of the bulk field to the two possible scaling dimensions $\Delta_\pm$ of the boundary primary field. As reviewed in Section \ref{singletsector}, in the integrable SYK models with unbroken higher-spin symmetry the scaling dimensions of the bilinear singlet operators $\mathcal{O}_n$ are $\Delta_n= n+1-\lambda$. The associated  bulk scalars $\varphi_n$ then have masses $m_n^2=(n-\lambda+1)(n-\lambda)$. It turns out that they form a higher-spin multiplet -- an irreducible representation of the algebra $\mathfrak{hs}[\lambda]$ -- and can be packaged into a zero-form $C$ valued in the twisted-adjoint representation of the higher-spin algebra \cite{Alkalaev:2019xuv}.\footnote{That the physical degrees of freedom can be described with the help of the twisted-adjoint representation is a fairly universal effect first identified in \cite{Vasiliev:1999ba}. However, there are notable exceptions; see, e.g. \cite{Skvortsov:2022syz}.} 

\paragraph{Twisted-adjoint representation interlude.} The algebra $\mathfrak{sl}(2,\mathbb{R})$ admits an involutive automorphism $\pi$ leaving invariant the Cartan subalgebra:
\begin{align}\label{pidef}
    \pi(e_{\pm})=-e_{\pm}\,,\qquad \pi(h)=h\,, \qquad \pi^2=\mathrm{id}\,.
\end{align}
This automorphism extends to $\mathfrak{gl}[\lambda]$ and $\mathfrak{hs}[\lambda]$, giving rise to the  {\it twisted-adjoint} representation $\rho$: 
\begin{equation}
a\mapsto \rho_a\,,\qquad \rho_a(x)=a\star x-x\star \pi(a)\,, \qquad \forall a,x \in\mathfrak{gl}[\lambda] \;(\mbox{or}\; \mathfrak{hs}[\lambda])\,.
\end{equation} 
The isomorphism $\mathfrak{sl}(2,\mathbb{R})\simeq \mathfrak{so}(1,2)$ allows for a geometric interpretation of the above relations in terms of the isometries of  two-dimensional anti-de Sitter spacetime $AdS_2$. Specifically, the basis elements $e_\pm$ are identified with the generators of transvections $P_a$ ($a=1,2$), while $h$ corresponds to the Lorentz boost $L_{ab}=\epsilon_{ab}L$. Consequently,  
\begin{align}\label{ads}
    [P_a,P_b]&= -L_{ab}\,, && [L_{ab},P_c]=P_a\eta_{bc}-P_b\eta_{ac}\,.
\end{align}
The twist flips the sign of the transvections, while leaving $L$ intact: $\pi(P_a)=-P_a$ and $\pi(L)=L$. The higher-spin algebra $\mathrm{Mat}[\lambda]$ is defined as the enveloping algebra of $P_a$ and $L$ subject to the relations induced by the quadratic Casimir operator.
Although $\mathfrak{gl}[\lambda]$ is a twisted-adjoint module for both $\mathfrak{gl}[\lambda]$ and $\mathfrak{hs}[\lambda]$, the subspace $\mathfrak{hs}[\lambda] \subset \mathfrak{gl}[\lambda]$ is not a submodule for either of these Lie algebras (because the unit element appears in twisted commutators).

The spectrum of matter fields is determined  by the decomposition of $\mathfrak{gl}[\lambda]$ under the twisted-adjoint action of $\mathfrak{sl}(2,\mathbb{R})$, which reads \cite{Alkalaev:2019xuv}\footnote{As a side remark, for $\lambda\in \mathbb{N}$, where (the quotient of) $\mathfrak{gl}[\lambda]$ reduces to $\mathfrak{gl}(n,\mathbb{R})$ the twist can be undone. Indeed, in the usual matrix realization of $\mathfrak{sl}(2,\mathbb{R})$, a $\mathfrak{gl}(2,\mathbb{R})$-matrix $J=\mathrm{diag}(1,-1)$ realizes the twist via $\bullet \rightarrow J\bullet J^{-1}$. This can be extended to its embedding into $\mathfrak{gl}(n,\mathbb{R})$. Therefore, the twisted-adjoint decomposition of $\mathfrak{gl}(n,\mathbb{R})$ coincides with the adjoint one, the latter given by the Clebsch-Gordon rule \eqref{tensorproduct}.}
\begin{align}\label{tadd}
    \mathfrak{gl}[\lambda]&= \bigoplus_{n=0}^{\infty} \mathcal{V}_{n+1-\lambda}\,.
\end{align}
This spectrum matches the spectrum $\Delta_n= n+1-\lambda$ discussed after \eqref{scalarfields}. Using \eqref{tensorproduct}, we observe that the twisted-adjoint decomposition (\ref{tadd})  coincides with the decomposition in irreducible $\mathfrak{sl}(2,\mathbb{R})$-modules  of the tensor square
\begin{align}
    \mathfrak{gl}[\lambda]&=\mathcal{V}_{\tfrac12(1-\lambda)}\otimes \mathcal{V}_{\tfrac12(1-\lambda)}\,.
\end{align}
As a result, the matter multiplet can be described by a zero-form $C$ valued in the twisted-adjoint representation (which is similar to other dimensions, see e.g. \cite{Vasiliev:1999ba}):
\begin{align}\label{DC=0}
    dC-\omega\star C+C\star \pi(\omega)&=0\,.
\end{align}
A general solution is $C=-g^{-1}\star C_0\star \pi(g)$, where $C_0$ is a constant element of $\mathrm{Mat}[\lambda]$. Currently, we have zero-forms $\phi$ and $C$, which take values in the adjoint and twisted-adjoint representations of the higher-spin algebra, respectively. Both of them can be realized within the adjoint representation of a larger algebra \cite{Alkalaev:2020kut}, which will be called the \textit{extended higher-spin algebra}. 

\subsection{Extended higher-spin algebra: unifying gravity and matter sectors} 

We extend $%
\mathrm{Mat}[\lambda]$ by the involutive automorphism $\pi$ passing to the smash-product algebra $
\mathrm{Mat}[\lambda]\rtimes \mathbb{Z}_2$, where $\mathbb{Z}_2=\{id,\pi\}$ is the group generated by $\pi$. More concretely, the latter associative algebra $
\mathrm{Mat}[\lambda]\rtimes \mathbb{Z}_2$ is defined as the extension of $
\mathrm{Mat}[\lambda]$ by a new generator $k$ subject to the relations 
\begin{equation}
    k\star k=1\,,\qquad k\star a\star k=\pi(a)\,,\qquad \forall a\in \text{Mat}[\lambda]\,.
    \end{equation}
A generic element $\mathbf{a}\in\mathrm{Mat}[\lambda]\rtimes \mathbb{Z}_2$ thus has the form $\mathbf{a}=a+a' \star k$, for $a, a'\in 
\mathrm{Mat}[\lambda]$, and the product of two elements is defined by the formula
\begin{align}
\mathbf{a}\star\mathbf{b}=(a+a'\star k)\star(b+b'\star k)&=\big(a\star b+a'\star \pi(b')\big) + \big(a\star b' +a'\star  \pi(b)\big) \star k\,.
\end{align}
The corresponding Lie algebra will be denoted $\mathfrak{gl}[\lambda]\rtimes\mathbb{Z}_2:=\mathrm{Lie}\big(\mathrm{Mat}[\lambda]\rtimes \mathbb{Z}_2\big)$.
It decomposes as the direct sum of its center $\mathfrak{u}(1)=\mathbb{R}\,1$ and a Lie algebra denoted $\mathfrak{ehs}[\lambda]$. In other words, $\mathfrak{gl}[\lambda]\rtimes\mathbb{Z}_2=\mathfrak{u}(1)\oplus\mathfrak{ehs}[\lambda]$, that is to say a generic element $\mathbf{a}\in\mathfrak{ehs}[\lambda]$ has the form $\mathbf{a}=a+a' \star k$ for $a\in
\mathfrak{hs}[\lambda]$ and $a'\in
\mathfrak{gl}[\lambda]$. 

The scalar field $\phi$ and the matter fields $C$ can be packaged into a single zero-form $\Cb$ valued in $
\mathrm{Mat}[\lambda]\rtimes \mathbb{Z}_2$: 
\begin{align}
    \Cb&= \phi + C \star k\,.
\end{align}
One can also extend $\omega$ to be a connection one-form $\omegab$ taking values in $
\mathrm{Mat}[\lambda]\rtimes \mathbb{Z}_2$, which leads to an additional one-form $\tilde \omega$ taking values in $
\mathrm{Mat}[\lambda]$:\footnote{Note that one can consistently truncate to $\phi\in\Omega^0(\mathcal{M})\otimes\mathfrak{hs}[\lambda]$ and  $\omega\in\Omega^1(\mathcal{M})\otimes\mathfrak{hs}[\lambda]$. However, one must keep $C\in\Omega^0(\mathcal{M})\otimes\mathfrak{gl}[\lambda]$ and $\tilde\omega\in\Omega^1(\mathcal{M})\otimes\mathfrak{gl}[\lambda]$.}
\begin{align}
    \omegab&= \omega + \tilde{\omega}\star k \,.
\end{align}
Setting $\Tr[a+a'\star k]=\Tr[a]$, we extend the trace from the original higher-spin algebra $\mathrm{Mat}[\lambda]$ to the extended higher-spin algebra $
\mathrm{Mat}[\lambda]\rtimes \mathbb{Z}_2$.\footnote{One needs to check that this still defines a trace operation. For the commutator of two generic elements as above we get $Tr\big[\,[a,b]_\star + a'\star \pi(b')- b'\star \pi(a')\,\big]$. An obvious property of the trace is the twist invariance, i.e., $\Tr[ \pi(f)]= \Tr [f]$ since the trace projects onto the unit $1\in 
\mathrm{Mat}[\lambda]$ and $\pi(1)=1$. As a result, the trace on the extended algebra does vanish on commutators.} The corresponding action \cite{Alkalaev:2020kut} is still of BF-type:
\begin{align}\label{BF}
    S_{EHS}[\Cb,\omegab]&= \Tr\int \Cb \star \Rb =\Tr\int \big[\phi \star R +C \star \pi(\tilde{R})\big]\,,
\end{align}
where $\Rb=d \omegab-\omegab\star \omegab=R+\tilde{R} \star k$, and we indicated the terms that contribute to the trace. 
In addition to the desired field equations, we get the covariant constancy condition for the one-form $\tilde{\omega}$: 
\begin{align}
    \tilde{R}=d\tilde{\omega}-\omega \star \tilde{\omega}-\tilde{\omega}\star \pi(\omega)=0\,.
\end{align}
Hence,  we can gauge it away unless the chosen boundary conditions forbid this. Then, the rest of the equations, i.e. \eqref{eoms} and \eqref{DC=0}, describe what has been planned: the topological (higher-spin) gravity multiplet of the flat connection one-form $\omega$ and the covariantly-constant zero-form $\phi$ in the adjoint representation, together with the matter multiplet of the covariantly-constant zero-form $C$ in the twisted-adjoint representation.
Equivalently, the adjoint representation of the extended higher-spin algebra $\mathfrak{ehs}[\lambda]$ decomposes into the sum of the adjoint representation and the twisted-adjoint representation 
of the higher-spin subalgebra $\mathfrak{hs}[\lambda]\subset\mathfrak{ehs}[\lambda]$.

\paragraph{Spectrum.}

Let us note that in three SYK models (free, nearly-free, double-scaled) the spectrum $\Delta_n$ of scaling dimensions for the singlet bilinear operators $\mathcal{O}_n$ spans the set $\mathbb N$ of all natural numbers (without zero for the $q=2$ model) for which something remarkable happens. Firstly, let us note that this integer spacing is consistent with their integrability (aka higher-spin symmetry). Secondly, the bulk dual of the tower of bilinears with all positive integer values is a tower of bulk scalar fields on $(A)dS_2$ with Dirichlet ($\Delta_+$) boundary condition that spans 
entirely the discrete series of unitary irreducible representations of the isometry group $SO(2,1)$. 
In other words, the twisted-adjoint representation of the algebra $\mathfrak{hs}[0]$ carries a unitary representation of the group $SO(2,1)$ which decomposes as the direct sum of all the discrete series of representations.

But there is more to say on this case. For definiteness, let us consider the nearly-free ($q=2$) SYK model.
The values $\Delta_n=n+1$ of the flavor-singlet bilinears $\mathcal{O}_n$ correspond the masses
$m_n^2=n(n-1)$ of the bulk scalar fields $\varphi_n$ on $AdS_2$.\footnote{These bulk fields are tachyonic for $dS_2$ but, nevertheless, unitary.} These are the ``shift-symmetric'' scalar fields (see e.g. \cite{Bonifacio:2018zex}). 
The Neumann boundary condition ($\Delta_-=1-\Delta_+$) on the same bulk scalar fields $\varphi_n$ correspond to shadow boundary fields $\widetilde{\mathcal{O}}_n$ with scaling dimensions $\widetilde{\Delta}_n=-n$. The bulk solutions with such Neumann conditions meet obstructions for the existence of the radial derivative of order $2n+1$ near the boundary. Assuming the vanishing of this obstruction gives a finite tail as radial expansion of such
unobstructed bulk solutions. These spaces of solutions span all the (non-unitary) finite-dimensional representations $\mathcal{D}_n$ of $SL(2,\mathbb{R})$, which correspond precisely to the spectrum of the BF theory for the higher-spin algebra. In this sense, the space of smooth solutions of the tower of bulk scalar fields $\varphi_n$ with masses $m_n^2=n(n-1)$ carries the sum of the adjoint representation (Neumann) and the twisted-adjoint (Dirichlet) representation 
of the higher-spin algebra $\mathfrak{hs}[0]$.
This feature is rather exceptional and might be hinting at some important feature of the extended higher-spin algebra $\mathfrak{ehs}[0]$.
As usual, another option is to allow for logarithms in the radial expansion in order to remove the obstruction. The bulk solutions for such Neumann conditions span the non-unitary Verma modules $\mathcal{V}_{-n}$ of $\mathfrak{sl}(2,\mathbb{R})$.
One may instead gauge away the finite-dimensional shift symmetries of these scalar fields, which would give back, on-shell, the discrete series of unitary irreducible representations.
Such fields have recently been studied in \cite{Farnsworth:2024yeh} where it was observed that they are conformal but are not usual $CFT_2$ since they do not have a gauge-invariant energy-momentum tensor.

\paragraph{BF/Poisson sigma model.} We see that the action and equations of motion are those of a Poisson sigma-model \cite{Ikeda:1993fh,Schaller:1994es} with a linear Poisson structure: 
\begin{align}\label{freeBF}
    S_{EHS}[\Cb,\omegab]&= \int \Cb^i d\omegab_i - \tfrac12 \pi^{ij}(\Cb) \, \omegab_i \wedge\omegab_j
\end{align}
(since any BF-model in two dimensions is such a Poisson sigma-model over the dual of the Lie algebra). Here, we introduced abstract indices $i,j, \ldots$ labeling the basis elements of the vector space $
\mathrm{Mat}[\lambda]\rtimes \mathbb{Z}_2$ and identified the dual space $
\big(\mathrm{Mat}[\lambda]\rtimes \mathbb{Z}_2\big)^\ast$ with $
\mathrm{Mat}[\lambda]\rtimes \mathbb{Z}_2$ by making use of the non-degenerate inner product $\langle a, b\rangle=\Tr [a\star b]$. The linear Poisson bivector $\pi^{ij}(\Cb)=f^{ij}_k \Cb^k$ is completely determined by the structure constants of the Lie algebra $\mathfrak{gl}[\lambda]\rtimes\mathbb{Z}_2$ associated to the associative algebra $\mathrm{Mat}[\lambda]\rtimes \mathbb{Z}_2$.

\paragraph{Another extended algebra and BF-model.}
For further purposes, let us mention another useful realization of the idea of an extended higher-spin algebra containing both the adjoint and twisted-adjoint representations. A close relative\footnote{A closely related algebra was first introduced  in \cite{Prokushkin:1998bq}, see also \cite{Sharapov:2024euk} for its application to $3d$ higher-spin gravities.} of the smash-product $
\mathrm{Mat}[\lambda]\rtimes\mathbb{Z}_2$ is the tensor product $
\mathrm{Mat}[\lambda]\otimes \mathrm{Mat}_2$. 

Let us interpret the associative algebra $\mathrm{Mat}_2$ of $2\times 2$ matrices as the Clifford algebra with two generators $\gamma_{1},\gamma_{2}$ and the relations
\begin{equation}
    \gamma_1\gamma_2+\gamma_2\gamma_1=0\,,\qquad  \gamma_1^2=\gamma_2^2=\mathrm{1}\,.
    \end{equation}
We redefine $P_a\rightarrow \tilde P_a=P_a\gamma_1$ and define $k=\gamma_2$. As a result, $k\tilde P_a k=-\tilde P_a$. Therefore, $
\mathrm{Mat}[\lambda]\rtimes\mathbb{Z}_2$ is the subalgebra of $\mathrm{Mat}[\lambda]\otimes\mathrm{Mat}_2$ spanned by elements of the form
\begin{align}
    f_0(\tilde P_a, L) + f_1(\tilde P_a,L)\,k \,.
\end{align}
The general elements of $\mathrm{Mat}[\lambda]\otimes\mathrm{Mat}_2$ can also depend on $\gamma_1$, which doubles the space. Note that conjugacy by $k=\gamma_2$ acts on $
\mathrm{Mat}[\lambda]\otimes\mathrm{Mat}_2$ explicitly as
\begin{align}
    k\,\Big(f_0(\tilde P_a, L,\gamma_1) + f_1(\tilde P_a,L,\gamma_1)k\Big)\,k= f_0(-\tilde P_a, L,-\gamma_1) + f_1(-\tilde P_a,L,-\gamma_1)\,k\,.
\end{align}
It is hard to make a clear preference between these two algebras ($
\mathrm{Mat}[\lambda]\rtimes\mathbb{Z}_2$ and $
\mathrm{Mat}[\lambda]\otimes\mathrm{Mat}_2$). Indeed, the spectrum of fields is determined by the equations linearized over the $AdS_2$ background  $\omegab_0=e^a\tilde P_a +\omega_0 L$. The equation $D_0 \Cb=0$ describes the Killing tensor fields and the matter fields. To show this, let us decompose $\Cb=\phi(\tilde P_a,L,\gamma_1) + C(\tilde P_a,L,\gamma_1)k$. The equations of motion are 
\begin{align}
    dC&= \omegab_0 \star C-C\star \pi(\omegab_0) \,, &d\phi&= \omegab_0 \star \phi-\phi\star \omegab_0\,.
\end{align}
Since the $AdS_2$ background $\omegab_0$ does not depend on $\gamma_1$ (explicitly), the only difference from $\mathrm{Mat}[\lambda]\rtimes\mathbb{Z}_2$ is the dependence on $\gamma_1$, which doubles the number of fields, but obeys the same individual field equations. 
Similarly to the case of $
\mathrm{Mat}[\lambda]\rtimes\mathbb{Z}_2$, we can write down the BF-model \eqref{freeBF} as an action for free higher-spin gravity based on $
\mathrm{Mat}[\lambda]\otimes\mathrm{Mat}_2$, which is also a Poisson sigma-model. The trace over $
\mathrm{Mat}[\lambda]\otimes\mathrm{Mat}_2$ is the usual matrix trace followed by the trace in $
\mathrm{Mat}[\lambda]$ (which was introduced in Section \ref{sec:HSJT}).

\paragraph{Further interactions?} While mathematically consistent, the BF-models above are physically incomplete since the matter fields (i.e. the scalars inside the zero-form $C$) should source the gravitational sector while the field equations above impose that the one-form connections $\omega$ and $\tilde\omega$ were necessarily flat on-shell. For example, the energy-momentum tensor $T_{mn}$ of scalar fields should contribute, e.g. in usual two-dimensional gravity:
\begin{align}\label{rhsTmm}
    R&= \epsilon_{ab}\, e^a \wedge e^b \,T\fud{m}{m}\,,
\end{align}
where $R$ is the component of the Cartan curvature two-form $T^a P_a +R\,L$ along the Lorentz generator $L$ and $T^a$ is the torsion two-form. The right-hand side of \eqref{rhsTmm} is the most general possibility in two dimensions. In the next section, we will introduce deformations of the above BF-theories that will include backreaction of matter on gravity.

\section{Two higher-spin gravity  models with backreaction}
\label{sec:FHSG}
By the end of the previous section we identified two relevant extended higher-spin algebras. Notably, such ambiguities do not arise in dimensions $D=4$ and higher, where massless higher-spin fields possess local physical degrees of freedom. Consequently, in $D \geqslant 4$, one-forms and zero-forms are inherently linked, and there appears to be no necessity for zero-forms in the adjoint representation (Killing tensors) or for one-forms in the twisted-adjoint representation (which lack a physical interpretation). However, in $D = 3$, these ambiguities become significant, and the spectrum can also be finite.\footnote{In three dimensions, it seems unavoidable to use the extension $
\mathrm{Mat}[\lambda]\otimes \mathrm{Mat}_2$, as $
\mathrm{Mat}[\lambda]$ itself does not support interactions; see \cite{Sharapov:2024euk} for more detail on this no-go theorem and on a yes-go result for $\mathrm{Mat}[\lambda]\otimes \mathrm{Mat}_2$.} 

We now construct two models, referred to as \textit{Model A} and \textit{Model B}, based on the two extended higher-spin algebras identified in the previous section; respectively:
\begin{itemize}
    \item[(A)] the smash product $\mathrm{Mat}[\lambda]\rtimes\mathbb{Z}_2$, and 
    \item[(B)] the tensor product $\mathrm{Mat}[\lambda]\otimes\mathrm{Mat}_2$.
\end{itemize}
The two models, however, have different physical status.

Model A is formulated at the level of formally consistent equations of motion. To extract concrete physical quantities -- such as correlation functions -- additional requirements, most notably locality, must be imposed.
 At a formal level, the problem reduces to constructing a deformation of the extended algebra $
\mathrm{Mat}[\lambda]\rtimes\mathbb{Z}_2$, which gives rise to a two-parameter family of algebras associated with $\mathfrak{sl}(2,\mathbb{R})$. Here one parameter is $\lambda$, while the second, $\nu$, corresponds to the deformation that owes its existence to $k$. 

Model B, by contrast, is more concrete: all interaction vertices are written explicitly as confi\-gu\-ration-space integrals. Moreover, its equations are variational, being derived from an action -- specifically, a Poisson sigma model.
At present, however, this model is restricted to $\mathrm{Mat}[\tfrac12]$. In fact, Model B is based on an oscillator realization of $\mathrm{Mat}[\lambda]$, which naturally selects $\lambda=1/2$. Nevertheless, it should be possible to extend the formulation to arbitrary values of $\lambda$.

\subsection{Model A}
\label{sec:FHSGA}

Before defining the model, it is useful to outline the general framework, which we refer to as formal higher-spin gravity.\footnote{The qualifier ``formal'' indicates that locality is not taken into account. Should an appropriate notion of locality in higher-spin gravity be defined (if it exists), one could always select representatives of the interaction vertices that satisfy it. This is precisely what occurs in Model B, where perturbatively local representatives of the vertices can be constructed.} This framework will also be relevant in the discussion of Model B.
Next, we define a specific deformation of $\mathrm{Mat}[\lambda]\rtimes\mathbb{Z}_2$, which is equivalent to constructing all formal interaction vertices. 

\subsubsection{Formal higher-spin gravities}
\label{sec:FHSGf}

To formulate equations of motion, we use the language of free differential algebras (FDA), which is closely related to $Q$-manifolds and $L_\infty$-algebras.\footnote{FDAs were first introduced by D. Sullivan in \cite{Sullivan77}, applied to supergravity in \cite{vanNieuwenhuizen:1982zf, DAuria:1980cmy} and to higher-spin interactions in \cite{Vasiliev:1988sa}. 
By “formal higher-spin gravity” we mean a mathematically rigorous formulation and generalization of the problem of deforming free equations for higher-spin fields in the form of FDAs, first studied in \cite{Vasiliev:1988sa,Vasiliev:1999ba} for specific examples.} FDAs provide a natural framework for deforming the free equations of motion discussed in Section \ref{sec:free}. The desired equations are sought in the form:
\begin{align}\label{FDA}
d\Phi&=\mathcal{V}_2(\Phi,\Phi) + \mathcal{V}_3(\Phi,\Phi,\Phi)+\mathcal{V}_4(\Phi,\Phi,\Phi,\Phi)+\mathcal{O}(\Phi^5)=Q(\Phi)\,,\qquad \Phi:=(\omegab,\Cb)\,.
\end{align}
More generally, one can take $\Phi$ to be a collection of differential forms of various degrees and interpret each inhomogeneous differential form $\Phi=\Phi(x, dx)$ as a smooth map from the total space of the shifted cotangent bundle $T^\ast[1]\mathcal{M}$ of a spacetime manifold $\mathcal{M}$ to a graded manifold $\mathcal{N}$ with coordinates $\Phi$. The target space $\mathcal{N}$ is equipped with a homological vector field $Q$. By definition, $Q$ has degree one and satisfies $Q^2=0$. This ensures the formal consistency and gauge invariance of the system  (\ref{FDA}), arising from the identity $d^2=0$.  The pair ($\mathcal{N}, Q$) is referred to as a $Q$-manifold. The pair $(\,T^\ast[1]\mathcal{M},d)$ made of the shifted cotangent bundle of the base manifold and of the exterior differential $d=dx^m\partial_m$ as homological vector field is also a $Q$-manifold. In these terms, each solution of the field equations (\ref{FDA}) defines and is defined by a morphism $\Phi:T^\ast[1]\mathcal{M} \to\mathcal{N}$ of $Q$-manifolds, which implies that $Q=\Phi_\ast (d)$. Expanding $Q$ near the point where $Q=0$ gives $L_\infty$ structure maps $\mathcal{V}_n$, while $Q^2=0$ reduces to certain bilinear relations $\sum \pm\mathcal{V}(..., \mathcal{V},...)=0$.  

In the higher-spin context, the free\footnote{What is meant by ``free'' here is somewhat different from the usual field theory approach, where ``free'' means no interactions and a fixed (simple) background such as a maximally symmetric space (being able to couple to an arbitrary gravitational background can already be considered as a form of interaction which is usually impossible for higher-spin fields). The flat connection $\omegab$ can be thought of as defining a maximally symmetric higher-spin background. The undeformed equation for the zero-form $\Cb$ describes the propagation of the higher-spin multiplet over any such flat background which is not necessarily AdS spacetime.} equations $d\Phi=\mathcal{V}_2(\Phi,\Phi)$ are determined by an associative algebra $\mathcal A$ alone:
\begin{align}\label{freesystemB}
d\omegab&=\omegab\star \omegab\,, &
d \Cb&=\omegab\star \Cb-\Cb\star \omegab\,.
\end{align}
The one-form $\omegab$ and the zero-form $\Cb$ take values in the (extended) higher-spin algebra $\mathcal A$. The free equations \eqref{freesystemB} are just the flatness condition for $\omegab$ and the covariant constancy of $\Cb$, both in the adjoint representation of the commutator Lie algebra $\mathfrak{g}=\mathrm{Lie}(\hsundeformed)$. Equivalently, they amount to  specifying a Lie algebra $\mathfrak{g}$ together with a representation $R$.\footnote{Indeed, \eqref{freesystemB} is just a particular case of $d\omegab=\tfrac12[\omegab,\omegab]$, $d\Cb=\rho_R(\omegab)\,\Cb$, where $[\bullet,\bullet]$ is a Lie bracket of $\mathfrak{g}$ and $\rho_R$ are the generators of some representation $R$ of $\mathfrak{g}$, where $\Cb$ takes values.}The crucial point, however, is that the non-linear deformation  relies on the fact that $\mathfrak{g}$ is the commutator Lie algebra of an associative algebra $\hsundeformed$. 

In \eqref{freesystemB} it is understood that the twist map is absorbed by transitioning to the extended higher-spin algebra $\hsundeformed=
\mathrm{Mat}[\lambda]\rtimes\mathbb{Z}_2$, as discussed in Section \ref{sec:HSJT+m}. The corresponding free equations \eqref{freesystemB} define the bilinear maps $\mathcal{V}_2(\Phi,\Phi)$. The deformation problem of the formal higher-spin gravity 
then reduces to finding the
higher structure maps $\mathcal{V}_{n\geqslant3}$ in 
\besubeqs\label{problem}
\begin{align}
d\omegab&=\omegab\star \omegab + \mathcal{V}_3(\omegab,\omegab,\Cb)+\mathcal{V}_4(\omegab,\omegab,\Cb,\Cb)+\mathcal{O}(\Cb^3)\,,\\
d\Cb&=\omegab\star \Cb-\Cb\star \pi(\omegab)+\mathcal{V}_3(\omegab,\Cb,\Cb)+\mathcal{O}(\Cb^3)\,,
\end{align}
\esubeqs
subject only to formal consistency, i.e. compatibility with $d^2=0$. At this stage, the right-hand sides of \eqref{problem} can be understood as a definition of the homological vector field $Q$. The target $Q$-manifold $\mathcal{N}$ is built on a graded vector space $V=V_{-1}\oplus V_0$ that is concentrated in two degrees, where each summand is isomorphic to the extended higher-spin algebra, $V_{-1}\simeq V_0 \simeq \hsundeformed$, and where $\omegab$, $\Cb$ take values in $V_{-1}$ and $V_0$, respectively. Equivalently, the multilinear maps on the right-hand sides (starting from the bilinear one) define an $L_\infty$-algebra or, more precisely, a Lie algebroid in this case. 

It is important to emphasize that formal consistency alone does not guarantee the physical consistency of the corresponding PDE system
\eqref{problem}.\footnote{Specifically, for higher-spin theories that are holographic duals of free/weakly coupled CFTs, the equations/actions are known to contain an infinite number of derivatives, which does not appear to align with the usual constraints of field theories, see, e.g., \cite{Bekaert:2015tva,Sleight:2017pcz,Ponomarev:2017qab}. However, for a more optimistic perspective in  Euclidean signature, see \cite{Neiman:2023orj}. It is also worth mentioning the model of Ref. \cite{Vasiliev:1995sv}, which is a formal higher-spin gravity in two dimensions. However, this theory is not relevant for the holographic duality with SYK models since it features a single scalar field in the bulk. }

The main observation  made in \cite{Sharapov:2019vyd} is that the following two problems (seemingly different) are closely related: 
\begin{itemize}
    \item[(a)] the problem of constructing interaction vertices $\mathcal{V}_{n\geqslant3}$, and 
    \item[(b)] the problem of deforming the underlying (extended) higher-spin algebra $\mathcal A$ as an associative algebra.
\end{itemize}
Problem (a) is apparently difficult, since it requires  constructing infinitely many multilinear maps that must  satisfy bilinear relations imposed by $Q^2=0$. In contrast, problem (b) is much simpler, since it concerns  associative algebra deformations. As shown in \cite{Sharapov:2019vyd} (see also below), given a deformation of the (extended) higher-spin algebra, one can construct an $A_\infty$-algebra\footnote{See Footnote \ref{Ainfty} for the definition of $A_\infty$-algebras, though the details will not be needed here} which, upon the graded-symmetrization (i.e. projection  to its associated $L_\infty$-algebra) provides a solution to  problem (a). Under certain technical assumptions -- for instance, allowing $\hsundeformed$ to be replaced by $\hsundeformed \otimes \mathrm{Mat}(n)$ -- it can be shown that all solutions of (a) arise in this way from solutions of (b).\footnote{The only known exception are the vertices of the chiral higher-spin theory in flat space, which do not come from deformations of the extended higher-spin algebra as described in \cite{Sharapov:2019vyd}. Otherwise, the argument relies on the fact that the commutator Lie algebra of $\hsundeformed \otimes \mathrm{Mat}(n)$ encodes the underlying associative structure, reducing the construction of the $L_\infty$-algebra to that of an $A_\infty$-algebra, and ultimately to deforming the associative algebra $\mathcal A$.}

Leaving the details to \cite{Sharapov:2019vyd}, let us assume that we are given a solution to the second, much simpler, problem (b). Specifically, we assume the existence of a one-parameter family of associative algebras $\hsdeformed$ that reduces to $\hsundeformed$ at $\nu=0$:
\begin{align}\label{fullproduct}
    a\ast b&= a\star b+\sum\limits_{k=1}^\infty \phi_k(a,b) \nu^k\,.
\end{align}
Then, the cubic and quartic vertices are obtained as
\begin{align}\label{firstV}
    \mathcal{V}_3(\omega,\omega,C)&= \phi_1(\omega,\omega)\star C\,,
\end{align}
\begin{align}\label{ffour}
    \mathcal{V}_4(\omega,\omega,C,C)&=\phi_2(\omega,\omega)\star C\star C +\phi_1\big(\phi_1(\omega,\omega),C\big)\star C\,,
\end{align}
with explicit formulas for higher vertices given in \cite{Sharapov:2019vyd}. This construction yields the most general form of interaction vertices, modulo field redefinitions (up to automorphisms at the $L_\infty$-algebra level).

Another result of \cite{Sharapov:2019vyd} is that the system is completely integrable: it admits a Lax pair, and the general solution can be written explicitly. The Lax pair is essentially the free system for $\hsdeformed$. Introducing a one-form $\omegabb$ and a zero-form $\Cbb$, one has
\begin{align}\label{Lax}
d\omegabb&=\omegabb\ast \omegabb\,,&
d \Cbb&=\omegabb\ast \Cbb-\Cbb\ast \omegabb\,.
\end{align}
Since the deformed product $\ast$ depends on $\nu$, the solutions to \eqref{Lax} inherit this dependence. The general solution of the nonlinear system \eqref{problem} then takes the form
\besubeqs\label{mauxmap}
\begin{align}
    \omega&=\omegabb+\omegabb' \ast \Cbb+\tfrac12 \omegabb''\ast \Cbb\ast \Cbb+\omegabb'\ast \Cbb'\ast \Cbb+(\omegabb'\ast' \Cbb)\ast \Cbb+\ldots \Big|_{\nu=0}\,,\\ 
    C&=\Cbb+\Cbb'\ast \Cbb+\ldots\Big|_{\nu=0}\,,
\end{align}
\esubeqs
where the prime $'$ denotes differentiation with respect to the ``spectral parameter'' $\nu$.  Moreover, the Lax pair solutions can be written in the ``pure gauge’’ form $\omegabb=-\gbb^{-1}\ast d\gbb$, $\Cbb= -\gbb^{-1}\ast \Cbb_0\ast \gbb$. Thus, solutions to the interacting system are expressed in terms of a one-parameter family of free solutions (for the deformed algebra $\hsdeformed$).
\footnote{This idea is reminiscent of earlier constructions such
the Vasiliev--Prokushkin integrating flow \cite{Prokushkin:1998bq}, Seiberg--Witten map \cite{Seiberg:1999vs}, and Nicolai map \cite{Nicolai:1980jc}.}

The vertices $\mathcal{V}_n$ define an $L_\infty$-algebra, which can also be understood as an infinite-dimensional Lie algebroid. However, it is not guaranteed that the equations have the form of a Poisson sigma-model. 
It is an open question whether there exists a representative where it is the case, so a genuine solution to the inverse variational problem is not known for Model A. Therefore, the interacting system, as it is, does not yet admit an action.

As an alternative, one can use the Lax pair \eqref{Lax}, which essentially encodes the same information. Assuming that $\hsdeformed$ admits a non-degenerate trace operation $\Tr$, one may write the Poisson sigma-model (in fact, BF-type) action
\begin{align}\label{EHS}
    S_{EHS}[\Cbb,\omegabb]&= \Tr\int \Cbb \ast (d\omegabb-\omegabb\ast \omegabb)\,,
\end{align}
whose equations of motion reproduce the Lax pair \eqref{Lax}. The $\nu=0$ component of this action reduces to the free action given in \eqref{BF} and \eqref{freeBF}.

To summarize: the construction of formal higher-spin gravities reduces to deforming the initial extended higher-spin algebra $\mathcal{A}=
\mathrm{Mat}[\lambda]\rtimes\mathbb{Z}_2$ as an associative algebra. This is a relatively simple problem, which we address in the next section. Importantly, the discussion above did not rely on any special property of $\mathcal{A}$, and therefore applies to any associative algebra $\mathcal{A}$. Thus, one obtains a broad class of nonlinear but integrable models associated with any one-parameter family of associative algebras $\mathcal{A}_\nu$.

\subsubsection{Deformation of the smash product}\label{sec:deformed}
In order to define a formal higher-spin gravity, we need to deform the associative algebra $${\mathcal{A}[\lambda]:=
\mathrm{Mat}[\lambda]\rtimes\mathbb{Z}_2}\,.$$ Let us first recall the commutation relations \eqref{comrelsl2} of $\mathfrak{sl}(2,\mathbb{R})\simeq \mathfrak{so}(2,1)$ in the Cartan--Weyl basis:
\begin{equation}\label{sl2}
    [h,e_{\pm}]=\pm e_{\pm}\,,\qquad [e_+,e_-]=h\,.
\end{equation}
This Lie algebra admits an involutive automorphism $\pi$ defined as follows: 
$$\pi(e_{\pm})=-e_{\pm}\,, \qquad\pi(h)=h\,.$$
The automorphism $\pi$  leaves invariant the Cartan subalgebra and allows us to extend the universal enveloping algebra $\mathcal{U}\big(\mathfrak{sl}(2,\mathbb{R})\big)$ by introducing an additional generator $k$. By definition, the new relations involving 
$k$ are:
\begin{equation}\label{kkk}
    k\,e_{\pm}=-e_{\pm}\,k\,,\qquad k\,h=h\,k\,,\qquad k^2=1\,,
\end{equation}
where the associative $\star$-product is left implicit.
This extension results in the smash-product algebra $\mathcal{U}\big(\mathfrak{sl}(2,\mathbb{R})\big)\rtimes \mathbb{Z}_2$. The quadratic Casimir element \eqref{C_2}
is invariant under the automorphism $\pi$. Hence, we can define the quotient algebra 
\begin{equation}
\mathcal{A}[\lambda]=\Big(\mathcal{U}\big(\mathfrak{sl}(2,\mathbb{R})\big)\rtimes \mathbb{Z}_2\Big)/\mathcal{J}_\lambda\,,
\end{equation}
where the two-sided ideal $\mathcal{J}_\lambda$ is generated by the central element
$C_2-\frac14(\lambda^2 -1)$, for some $\lambda\in \mathbb{R}$. 

The central observation is that the associative algebra $\mathcal{A}[\lambda]$ admits a one-parameter deformation $\mathcal{A}[\lambda,\nu]$ defined by the relations \eqref{kkk} together with:
\begin{equation}\label{hee}
\begin{array}{c}
[h,e_{\pm}]=\pm e_{\pm}\,,\qquad [e_+,e_-]=(1+\nu k)\,h\,,\\ [5mm] 
\displaystyle e_+e_-+e_-e_+ +h^2-\frac14(\lambda^2 -1)-\frac{\nu}{2}\,k+\frac{\nu^2}{4}=0\,.
\end{array}
\end{equation}
For $\nu=0$, we clearly get the original algebra, i.e. $\mathcal{A}[\lambda]=\mathcal{A}[\lambda,0]$.
Let $\mathcal{B}_\nu$ be the associative algebra defined by \eqref{kkk} and the first line in \eqref{hee}.
One can verify that the polynomial element $\widetilde{C}_2(\nu):=e_+e_-+e_-e_+ +h^2-\frac{\nu}{2}k$ of degree two in the 4 generators 
$e_+$, $h$, $e_-$, $k$
is an element in the centre $\mathcal{Z}_\nu$ of the algebra $\mathcal{B}_\nu$. The deformed algebra $\mathcal{A}[\lambda,\nu]$ is nothing but the quotient of the algebra $\mathcal{B}_\nu$ by the ideal generated by $\widetilde{C}(\nu)-\frac14(\lambda^2 -1)+\frac{\nu^2}{4}$. 
As shown in Appendix \ref{app:trace}, there exists a non-degenerate trace over this algebra $\mathcal{A}[\lambda,\nu]$. This is enough to show the existence of the corresponding BF-action \eqref{EHS} for the Lax pair.

A generic element of the deformed algebra $\mathcal{A}[\lambda,\nu]$ is represented by a function $f(e_+, h, e_-, k)$, once a certain ordering of the generators has been fixed, and $f$ is then simplified by using \eqref{kkk} and \eqref{hee}. For example, $e_+e_-$ and $e_-e_+$ can always be eliminated (see Appendix \ref{app:trace} for more details). Using relations (\ref{kkk}) and (\ref{hee}), one can, in principle, compute the product $f \ast g$, expand it in powers of $\nu$, and extract the functions $\phi_k(\bullet,\bullet)$ that appear in (\ref{fullproduct}). As explained above, the functions $\phi_k$ allow one to construct the interaction vertices in formal higher-spin gravity.

Another natural basis for the algebra is the AdS-basis. In this basis, the defining relations take the form
\begin{align}\label{deformedJoseph}
    [L,P_a]&= \epsilon_{ab}P^b \,,& [P_a,P_b]&= -\epsilon_{ab}(1+\nu k )L \,,& kP_ak=-P_a \,, && kLk=L \,,&& k^2=1\,,
\end{align}
and the Casimir operator determines the extra relation $$P_aP^a+ L^2 -\frac14(\lambda^2 -1)-\frac{\nu}{2}k+\frac{\nu^2}{4}=0\,.$$ 

There is also a $d$-dimensional perspective on the same algebra. $\mathrm{Mat}[\lambda]$ is the $d=1$ case of the algebra $\mathcal{A}_E$ of higher symmetries of the conformal Laplacian \cite{Eastwood:2002su}. It is defined as a quotient of $\mathcal{U}(\mathfrak{so}(d,2))$ by a two-sided ideal, known as the Joseph ideal. The Joseph ideal for $d>1$ is generated by more elements than just $C_2-c$. Formally, it is the maximal ideal such that the quotient algebra remains infinite-dimensional. One can still define the smash product of $\mathcal{A}_E$ with $\mathbb{Z}_2$ and deform it along $k$ \cite{Sharapov:2019pdu}. The $d$-dimensional counterpart of $\mathcal{A}[\lambda,\nu]$ is then given by the quotient of the deformed algebra $\mathcal{U}\big(\mathfrak{so}(d,2)\big)\rtimes \mathbb{Z}_2$ by  the deformed Joseph ideal.

\subsection{Model B}
\label{sec:FHSGB}
At the level of equations of motion, Model B employs the same formalism as described in Section \ref{sec:FHSGf}.  However, interaction vertices in this model exhibit more structure. Two key properties are worth noting: first, the vertices are explicitly local, and second, the equations take the form of a Poisson sigma-model. In its current form, the model is explicitly known for $\lambda=1/2$ where an oscillator realization is particularly helpful.  Nevertheless, the model should generalize to arbitrary values of  $\lambda$.

\paragraph{Oscillator realization.} There is an alternative realization of the algebra $
\mathrm{Mat}[\lambda]$ for $\lambda=1/2$ in terms of oscillators.  Let us define (indices $A,B,\ldots=1,2$ are the ones of the fundamental representation of $\mathfrak{sl}(2,\mathbb{R})$)
\begin{align}
    [y_A,y_B]&= -2\epsilon_{AB}\,,
\end{align}
i.e. $[y_2,y_1]=2$ and we can choose $y_2=\sqrt{2}a$, $y_1=\sqrt{2}a^\dag$ in terms of the usual annihilation/creation operators. Recall that quadratic polynomials
\begin{align}
    t_{AB}&=-\tfrac14 \{y_A,y_B\}\,,
\end{align}
generate $\mathfrak{sl}(2,\mathbb{R})$ through the commutators \eqref{osccomA}. The Casimir is fixed as $C_2=-3/16$, which corresponds to $\lambda=1/2$. This shows that $\mathrm{Mat}[\frac12]$ is isomorphic to the subalgebra $A^{\text{even}}_1\subset A_1$ of the  Weyl algebra that is spanned by even polynomials.

\subsubsection{Vertices}

To begin with, let us fix the Weyl algebra as $[y_A,y_B]_\star=-2\hbar\,\epsilon_{AB}$, where $\hbar$ is a free parameter of the construction (for ease of comparison with \cite{Sharapov:2022wpz,Sharapov:2022nps} we keep $\hbar$ explicit, but one could have set $\hbar=1$ in the present paper). The algebraic structure we are going to describe is an $A_\infty$-algebra\footnote{\label{Ainfty}An $A_\infty$ algebra $\mathbb{A}$ is a graded space $V$ together with multilinear maps $T^n V\rightarrow V$, $m_n(\bullet,...,\bullet)$ for $n=1,2,...$ of total degree $1$ such that the Stasheff relations are satisfied, which are schematically 
\begin{align}
  \sum_{i+j=n}\sum_k (\pm)m_i\big(v_1,....,v_k, m_j(v_{k+1},...,v_{k+j-1}), v_{k+j},...,v_{i+j}\big)&=0\,,\qquad v_i\in V.
\end{align} 
The sign is determined by the Koszul rule. In our case, $\mathbb{A}$ is minimal, so $m_1(\bullet)=0$. 
} denoted $\mathbb{A}$, whose graded space is $V=V_0 \oplus V_{1}$, with $V_1\simeq A$ and $V_0\simeq A^*$ for some associative algebra $A$, i.e. it is built on an associative algebra $A$ and its dual bimodule $A^*$. Given any $A_\infty$-algebra $\mathbb{A}$, one can tensor it with any associative algebra $B$ to get a new $A_\infty$-algebra $\mathbb{A}\otimes B$. Since the $A_\infty$-maps and relations do not have/impose any symmetry on the arguments, one can always assume that arguments $a_1, a_2,\ldots, a_n \in \mathbb{A}\otimes B$ are replaced by elements from $V\otimes B$. In our case, $B=\mathrm{Mat}_2$ and matrices are multiplied in the same order as the arguments. Below, we borrow the $A_\infty$-algebra $\mathbb{A}$  of chiral higher-spin gravity in four dimensions \cite{Sharapov:2022wpz,Sharapov:2022nps}. The same $A_\infty$-algebra also describes dynamics in two dimensions, as will be explained below. For chiral higher-spin gravity, the associative algebra $B$ is another copy of the Weyl algebra.

More formally, each $A_\infty$-map of $\mathbb{A}\otimes B$, which we denote $M_n(a_1,\ldots ,a_n)$, has the following factorized form
\begin{align}
    M_n(a_1,\ldots ,a_n)&= m_n(v_1,\ldots , v_n) \otimes (b_1\circ\cdots  \circ b_n)  \,,
\end{align}
where $a_i=v_i \otimes b_i$ with $v_i\in V$ and $b_i\in B$, while $m_n$ are the structure maps of $\mathbb{A}$. The product in $B$ is denoted $\circ$, but since $\circ$ is just the matrix product, we will omit this notation from now on. Another restriction is that we have $\lambda=1/2$, so the higher-spin algebra is isomorphic to the even subalgebra of the  Weyl algebra, $A=\mathrm{Mat}[\frac12]\simeq A^{\text{even}}_1$, for which the realization via the Moyal-Weyl star-product will be used.

\begin{figure}
    \centering
    \begin{tikzpicture}
\draw[thin, gray] (0,0)--(0,4)--(4,4)--(4,0)--cycle;
\draw[thin, gray] (0,0) -- (4,4);
\fill[black!7!white] (0,0) -- (4,0)--(4,4)--(3.5,2)--(2.5, 0.7) --(1.5,0.2)-- cycle;

\draw[-latex, thin, gray] (0,0)--(0,5);
\draw[-latex,thin,gray](0,0)--(5,0);

\draw[latex-, thick, color=blue ] (0,0)--(4,0);
\draw[latex-, thick, color=blue ] (4,0)--(4,4);
\draw[-latex, thin, color=red ] (0,0)--(1.5,0.2);
\draw[-latex, thin, color=red ] (1.5,0.2)--(2.5,0.7);
\draw[-latex, thin, color=red ] (2.5,0.7)--(3.5, 2);
\draw[-latex, thin, color=red ] (3.5, 2) -- (4,4);

\coordinate [label=right:${\vec{q}_a}$] (B) at (4,2);
\coordinate [label=below:${\vec{q}_b}$] (B) at (2,0);

\coordinate [label=below:$_0$] (B) at (0,0);
\coordinate [label=below:$_{1}$] (B) at (4,0);
\coordinate [label=left:$_{1}$] (B) at (0,4);

\coordinate [label=left:${u}$] (B) at (0,4.5);
\coordinate [label=below:${v}$] (B) at (4.5,0);

\coordinate [label=above:${\vec{q}_1}$] (B) at (0.9,0.1);
\coordinate [label=above:${\vec{q}_2}$] (B) at (1.8,0.4);
\coordinate [label=above:${\vec{q}_3}$] (B) at (2.7,1.2);
\coordinate [label=left:${\vec{q}_4}$] (B) at (3.7,2.8);

\coordinate [label=left:$_+$] (B) at (3.7, 0.7);

\end{tikzpicture}

\caption{The configuration space can be realized as the set of maximally concave polygons in a canonical form, where two vectors at the convex angles are aligned with the edges of the unit square. Any maximally concave polygon can be brought to this form by   an affine transformation. }
\label{ST}
\end{figure}
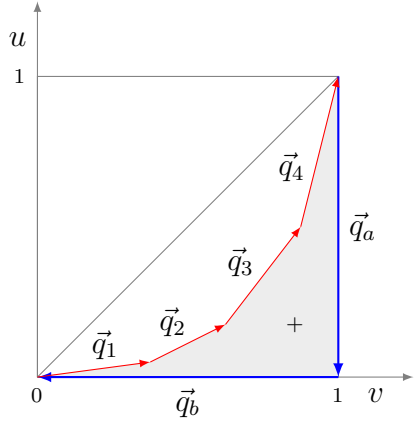

We identify elements from $V_0$ and $V_1$ with even functions $f(y)=f(-y)$ of the commuting variables $y_A$. The structure maps $m_n$ can be represented by poly-differential operators \cite{Sharapov:2022wpz}. To give examples of $m_n$ it is convenient to introduce auxiliary variables $y^A_i$ and set $p_i^A\equiv \pl^A_{y_i}\equiv \pl/ \pl y^A_i$. Then, $m_n$ can be represented as
\begin{align}
    m_n\big(f_1(y),...,f_n(y)\big)&= \mathcal{O}_n(y, p_1,\ldots ,p_n)\,f_1(y_1)\cdots f_n(y_n)\Big|_{y_i=0}\,.
\end{align}
In addition, we set $p_0\equiv y$ to make the notation more uniform and $p_{ij} \equiv -p^A_i p^B_j \epsilon_{AB}$. For example, $\exp[y\cdot p_1]f(y_1)=f(y_1+y)$ is the translation operator. Now, the Moyal--Weyl star-product defines the first structure map:
\begin{align}\label{hsalgebra}
    m_2(f_1,f_2)&= \exp{[p_{01}+p_{02}+\hbar\, p_{12}]}f_1({y}_1)\, f_2({y}_2)\Big|_{{y}_{1,2}=0} \equiv (f_1\star f_2)(y)\,,
\end{align}
where $f_{1,2}\in V_1$. In order to define the bimodule structure on $A^*$ we choose a non-degenerate pairing in the form
\begin{align}
    \left\langle f, g\right\rangle=\exp\left[p_1\cdot p_2\right]f\left(y_1\right){g\left(y_2\right)}\big|_{y_{1,2}=0}\,.
\end{align}
The bimodule structure gives two more maps via $\langle \bullet, \bullet \rangle$:
\begin{equation}
    \begin{split}
        &m_2(\omega,C)=+\exp{[\hbar\, p_{01}+ p_{02}+p_{12}]}\, \omega({y}_1)\, C({y}_2)\Big|_{\bar{y}_i=0}\,,\\
        &m_2(C,\omega)=-\exp{[p_{01}-\hbar\, p_{02}-p_{12}]}\, C({y}_1)\, \omega({y}_2)\Big|_{{y}_i=0}\,.
    \end{split}
\end{equation}
The bilinear vertices are defined as 
\begin{align*}\mathcal{V}_2(\omegab,\omegab)&=M_2(\omegab,\omegab)=\omegab\star \omegab\,, & \mathcal{V}_2(\omegab,\Cb)&=M_2(\omegab,\Cb)=+\,\omegab\star \Cb\,, \\
    \mathcal{V}_2(\Cb,\omegab)&=M_2(\Cb,\omegab)=-\,\Cb\star \omegab\,, &
    \mathcal{V}_2(\Cb,\Cb)&=0\,,
\end{align*} 
where the star-product representations of $M_2(\omegab,\Cb)$ and $M_2(\Cb,\omegab)$ are true for $\hbar=1$ and for even functions of $y_A$. We will not list all the structure maps explicitly (see \cite{Sharapov:2022wpz}), but for  an all-order example, we have
\begin{align}
    m_{n+2}(\omega,\omega, C,\ldots, C)= p_{ab}^n \int \exp\Big[ (1-\sum_i u_i) p_{0a} +(1-\sum_i v_i) p_{0b} +\sum_i u_i p_{a,i}+\sum_i v_i p_{b,i}+ \notag\\
      +\hbar\, \Big(1+\sum_i (u_i-v_i) +\sum_{i,j} u_iv_j \sign(j-i) \Big ) p_{ab} \Big]\,\omega(y_a)\omega(y_b) C(y_1)\cdots C(y_n)\Big|_{y_{a,b,i}=0}\,.\label{allorder}
\end{align}
The integration is over a compact domain in $\mathbb{R}^{2n}$ parametrized by the $u_i$'s and $v_i$'s. Other structure maps have a very similar form, which only differs by certain permutations of the integration parameters. The compact domain is defined as follows. The vectors $\vec{q}_i=(u_i,v_i)\in\mathbb{R}^2$ form a closed (maximally) concave polygon when supplemented with two more vectors $\vec{q}_a=(-1,0)$ and $\vec{q}_{b}=(0,-1)$, as illustrated in Fig. \ref{ST}. The $\hbar$-term in the exponent is just twice the area of this polygon times $p_{ab}$. 

\subsubsection{Poisson sigma-model}

One important property of the structure maps is that they define a Poisson structure on the space coordinatized by $V_0$\,, i.e., by the zero-form $\Cb(y)$. Introducing (once again) the abstract indices $i,j,k,...$ for the basis in $V_0$ and $V_1$, the equations of motion read
\begin{align}\label{PSM}
    d\omegab_k=\tfrac12\partial_k\pi^{ij}(\Cb)\,\omegab_i\,\omegab_j\,, \qquad  \qquad  d\Cb^i=\pi^{ij}(\Cb)\,\omegab_j\,.
\end{align}
Therefore, the action of Model B has the form of a Poisson sigma-model with highly nontrivial corrections that deform the linear Poisson structure $\pi^{ij}(\Cb)=f^{ij}_k \Cb^k$ that originates from $\mathrm{Lie}(A)$, where $A=\mathrm{Mat}[\tfrac12]\simeq A^{\text{even}}_1$.  

The fact that the equations have the form of a Poisson sigma-model can be seen at the level of the underlying $A_\infty$-algebra. The fact that $\mathbb{A}$ is built on $A$ and $A^*$ implies that there is a canonical pairing between all structure maps and elements of $V$. This allows one to impose the cyclicity on the structure maps. Such an $A_\infty$-algebra is called a pre-Calabi--Yau algebra, cf. \cite{kontsevich2021pre,IYUDU202163,Kontsevuch:2006jb}. This property can also be understood as a non-commutative analog of the Poisson structure. Indeed, the projection of a pre-Calabi-Yau $A_\infty$-algebra $\mathbb{A}$ via the symmetrization map to an $L_\infty$-algebra gives a Poisson structure (to be precise it gives a specific type of an $L_\infty$-algebra that is the Lie algebroid corresponding to the equations of a Poisson sigma-model, from which the Poisson structure can be read off). In addition, any pre-Calabi-Yau algebra can be tensored with a matrix algebra to get another pre-Calabi-Yau algebra, which is the property we used to add the factor $B=\mathrm{Mat}_2$. 

In a bit more detail, let us define the \textit{master action} \begin{equation}S=\tfrac12 \pi^{ij}(C) \,\omega_i\,\omega_j\,.\end{equation} 
The nilpotency condition $Q^2=0$ underlying \eqref{PSM} can be understood as the \textit{master equation}
\begin{equation}\{S,S\}= \frac{\pl S}{\pl \Cb^i}\frac{\pl S}{\pl \omegab_i}=0\end{equation}
with respect to the odd symplectic structure $\Omega=d\Cb^i \wedge d\omegab_i$. In our case, the homological vector field is
\begin{align}
    Q&=\tfrac12\partial_k\pi^{ij}(\Cb)\,\omegab_i\,\omegab_j \frac{\pl}{\pl \omegab_k} + \pi^{ij}(\Cb)\,\omegab_j\frac{\pl}{\pl \Cb^i}\,.
\end{align}
More generally, $Q$ can be expanded as follows:
\begin{align}
    Q&=\sum_n\mathcal{V}_n(\Phi,\ldots,\Phi)\frac{\pl}{\pl \Phi}\,.
\end{align}
The Taylor expansion of the nilpotency condition $Q^2=0$ gives the $L_\infty$-relations on the maps $\mathcal{V}_n$. Likewise, if we assume that $\Phi$ are graded but non-commutative coordinates, define
\begin{align}
    \mathbb{Q}&=\sum_n M_n(\Phi,\ldots,\Phi)\frac{\pl}{\pl \Phi}
\end{align}
and assume that (noncommutative) derivatives ${\pl}/{\pl \Phi}$ respect the graded Leibnitz law, then the Taylor expansion of $\mathbb{Q}^2=0$ produces the $A_\infty$-relations for $M_n$. Lastly, let us define a (noncommutative) master action
\begin{align}
    \mathbb{S}&= \sum \langle M_n(\Cb,\ldots,\Cb,\omegab,\Cb,\ldots,\Cb)| \omegab\rangle\,.
\end{align}
Then, the $A_\infty$-relations together with the property that the structure maps are cyclic-invariant with respect to the pairing $\langle\bullet|\bullet\rangle$ can be encoded in the non-commutative master-equation \begin{equation}[\mathbb{S},\mathbb{S}]_{\text{necl}}=0\,,\end{equation} where $[\bullet,\bullet]_{\text{necl}}$ is called the \textit{necklace bracket}. It is a non-commutative analog of the Schouten--Nijenhuis bracket and can be defined with the help of the non-commutative symplectic structure $\Omega=d\Cb^i \wedge d\omegab_i$. 

To summarize, every $A_\infty$-structure gives an $L_\infty$-structure via the symmetrization map. A particular class of $A_\infty$-algebras are based on an associative algebra $A$ and its dual bimodule $A^*$. With the help of the natural pairing one can require the algebra to be cyclic. It turns out that the corresponding $L_\infty$-projection is nothing but a Poisson structure (the associated $Q$ gives the equations of motion of a Poisson sigma-model, in other words, it describes a Poisson Lie algebroid). Model B is an example of this construction for $A=\mathrm{Mat}[\tfrac12]\otimes \mathrm{Mat}_2$. It is a perturbatively local field theory in $(A)dS_2$ which admits an action principle in the form of a Poisson sigma-model. 

\section{Conclusion}
\label{sec:discussion}

Various higher-spin extensions of JT gravity have been proposed  in \cite{Alkalaev:2013fsa,Grumiller:2013swa,Alkalaev:2014qpa,Alkalaev:2020kut}. One of the ideas is to replace $\mathfrak{sl}(2,\mathbb{R})$ with a higher-spin algebra (e.g. $\mathfrak{hs}[\lambda]$ in \cite{Alkalaev:2014qpa} or $\mathfrak{sl}(n,\mathbb{R})$ in \cite{Alkalaev:2013fsa,Grumiller:2013swa})\footnote{A somewhat related model (extension of JT gravity) is studied in \cite{Chirco:2024ubu}. } in the BF formulation of JT gravity. A peculiar feature of such extensions is that matter fields can be incorporated naturally by taking an extended higher-spin algebra $\mathfrak{gl}[\lambda]\rtimes\mathbb{Z}_2$, which contains the twisted-adjoint module of the original higher-spin algebra alongside the adjoint one, cf. \cite{Alkalaev:2020kut}.

In this paper, we introduced two models, A and B, that deform the BF-model of \cite{Alkalaev:2020kut} and have non-trivial interactions between the matter and higher-spin sectors. The existence of these models, that incorporate both physical and auxiliary fields, reflects the presence of two closely related extended higher-spin algebras. 

\paragraph{Model A vs B.} The formalism we employ allows one to construct vertices directly at the level of the equations of motion. Remarkably, the equations of Model B already come in a form that ensures the existence of a Lagrangian that has the form of a Poisson sigma-model. It is less clear whether the equations of motion for Model A can be cast into a similar form. Nevertheless, a sensible BF-model (a particular case of Poisson sigma-model) can be formulated for the {\it deformed} extended higher-spin algebra, whose equations of motion define a Lax pair for Model A.

\paragraph{One-parameter family for Model A.} In Section \ref{sec:deformed}, we constructed the deformed extended higher-spin algebra underlying Model A. It is a simple associative algebra, which is a deformation of $\mathrm{Mat}[\lambda]$ extended by a reflection automorphism. The algebra is specified by the simple quadratic relations \eqref{MatlambdaZ2deformed1}-\eqref{MatlambdaZ2deformed3} that deform the commutation relations \eqref{undefalg} and specify the value of the Casimir element. However, despite its simplicity, the algebra does not seem to be known in the literature. It is a natural playground for deformations of the basic commutation relations in quantum mechanics. As shown in Appendix \ref{app:trace}, this algebra admits a non-degenerate trace, thereby allowing the existence of a BF-model for the Lax pair of Model A.

\vspace{2mm}
Some possible directions of future  investigations are:

\paragraph{One-parameter family for Model B.} The higher-spin algebra underlying Model B has $\lambda=1/2$, which is due to the fact that it admits a simple oscillator realization that was used for chiral higher-spin gravity. Mathematically, $\lambda=1/2$ is a generic point in the one-parameter family and it should be possible to define a pre-Calabi-Yau $A_\infty$-algebra (a non-commutative analog of a Poisson structure) that encodes the equations of motion and the action for any generic $\lambda$. Therefore, we expect that for any $\lambda$ there exists an interacting higher-spin theory in the form of a Poisson sigma-model with the required features for holographic duality with SYK-type models -- such as non-trivial interactions between matter and higher-spin gauge fields.   We hope to address this problem in future work.

\paragraph{Asymptotic symmetries and Schwarzian actions.} The main focus of the present paper is the search for a bulk theory whose matter sector could reproduce an SYK-type tower of singlet operators, rather than the usual JT/SYK duality between JT gravity and the Schwarzian action. In fact, two issues which have not been addressed here are the asymptotic symmetries and the Schwarzian sector of our proposed models. This would require supplementing the action with an appropriate boundary term, presumably a particle-like action on the corresponding group manifold (i.e. the higher-spin group here). In Section \ref{sec:symSYK}, we mentioned the existence of a higher-spin analogue of the Schwarzian action, corresponding to the infinite-dimensional higher-spin algebra $\mathfrak{gl}[\lambda]$, in the two-body case ($q=2$).
It would be interesting to generalize the analysis on the asymptotic symmetries \cite{Grumiller:2013swa} and Schwarzian actions \cite{Gonzalez:2018enk} of higher-spin gravity for $\mathfrak{sl}(n,\mathbb{R})$ to the infinite-dimensional higher-spin algebras $\mathfrak{hs}[\lambda]$ and $\mathfrak{ehs}[\lambda]$.

\paragraph{Correlation functions of SYK-type holograms for Model A.}
Since locality is not yet under control for Model A, it is a challenge to compute correlation functions beyond the leading order. Nevertheless, one can try to apply the ideas of higher-spin invariant observables \cite{Sezgin:2011hq,Colombo:2012jx,Didenko:2012tv,Bonezzi:2017vha}, with complete classification obtained in \cite{Sharapov:2020quq}. In particular, there is a simple class of observables constructed via the trace $\Tr(a_1\ast \ldots \ast a_n)$ that gives correlation functions of the free CFTs \cite{Colombo:2012jx,Didenko:2012tv,Bonezzi:2017vha,Scalea:2023dpw}. It would also be interesting to identify the deformed extended higher-spin algebra of Section \ref{sec:deformed} directly on the boundary, where it can serve as a spectrum-generating algebra for the scaling dimensions. 

\paragraph{Correlation functions of SYK-type holograms for Model B.} Model B is a well-defined perturbatively local field theory with an action in the form of a Poisson sigma-model. Therefore, it is immediately suitable for holographic calculations of correlation functions via the Gubser--Klebanov--Polyakov--Witten prescription by computing the on-shell action. One puzzle, however, is that this ``easily accessible'' point $\lambda=1/2$ -- instrumental for Model B -- does not correspond to any of the higher-spin symmetry enhanced SYK points in Table \ref{table2}. Therefore, it seems that the $\lambda$-family of Poisson sigma-models alluded to above would define a line of bulk theories that may have several ``intersections'' with the hypothetical bulk duals of SYK models with higher-spin symmetry enhancement. One appealing feature of this one-parameter family of higher-spin bulk theories is that they should be perturbatively local and (at least, naively) unitary. Furthermore, they would admit a simple action -- a Poisson sigma-model. Let us repeat that these Poisson sigma-models are non-topological (which is possible because the target space is infinite-dimensional) and, in fact, the tower of scalar matter fields defines the propagating degrees of freedom. 

\paragraph{Consistent truncations.} Strictly speaking, the spectrum of Model B is doubled compared to the desired SYK models (with a single tower of flavor singlets) but there is a proposal that it can be truncated consistently \cite{Bonezzi:2015igv}. More generally, it could be interesting to investigate the existence of consistent truncations for Models A and B, possibly leading to finite spectra.

\paragraph{First-order formulation.} It may be instructive to eliminate the auxiliary fields present in the first-order formulation using the standard techniques of free differential algebras  and pass from a first order (``frame-like'') to a second order (``metric-like'') approach, see e.g. \cite{Fredenhagen:2014oua} for a similar story in three dimensions. This could also clarify the concept of ``higher-spin geometry'', at least in two dimensions, by considering standard geometric experiments, such as point particle probes \cite{Ivanovskiy:2025kok,Ivanovskiy:2025ial}. In particular, it would be interesting to look for a definition of higher-spin ``black holes'' in two dimensions along the lines of what is done for usual dilaton gravity \cite{Grumiller:2002nm}.

\section*{Acknowledgments}
\label{sec:Aknowledgements}
We would like to thank M.~Pannier, and especially K.~Alkalaev, for exchanges about higher-spin gravity in two dimensions, as well as C.~Peng for discussions about SYK model. 
The work of E. S. was partially supported by the European Research Council (ERC) under the European Union’s Horizon 2020 research and innovation programme (grant agreement No 101002551). 

\appendix

\section{Double scaling limit of SYK model as two-dimensional Liouville-like theory}\label{app:Liouville}

This section presents a slight generalisation (including a source) of the original derivation (in Appendix B of \cite{Cotler:2016fpe}) of a Liouville-like theory from the double scaling limit of SYK model.

Consider the $q$-body SYK model with first-order kinetic term, i.e. $\lambda=1$ and $\hat{K}_1=i\partial_\tau$. 
Let $\hat{G}_0=(i\partial_\tau)^{-1}$ be the free propagator. Its integral kernel reads $G_0(\tau)=-\,i\,\text{sgn}(\tau)$. Note that it obeys the property $G_0(\tau)^2=-1$ which will be used later on. 

For Majorana fields $\vec\chi$, let us assume the bilocal fields $G$, $\Sigma$ and $H$ to be pure imaginary, e.g. \begin{equation}
    H(\tau_1,\tau_2)=i A(\tau_1,\tau_2)\,,\quad \text{with}\quad A(\tau_2,\tau_1)=-A(\tau_1,\tau_2)\,.
\end{equation} 
The collective field theory \eqref{collectiveFT} can be written as
\begin{equation}\label{collect}
   \,I[A,G,\Sigma\,;N,J,q]\,=\,\,N\,\Bigg(-\text{Tr}\Big[\,\log\Big(\,i(\hat \partial_\tau+\hat A)-\hat\Sigma\Big)\,+\,\hat\Sigma\,\hat G\,\Big]
    +\,\frac{J^2}{q^2}\int d\tau_1 \, d\tau_2\,G(\tau_1,\tau_2)^q\Bigg)\,.
\end{equation}

Consider the following field redefinitions 
\begin{equation}\label{fieldredefs}
    G(\tau_1,\tau_2)=G_0(\tau_1,\tau_2)\,\big[1+\frac1{q}\,g(\tau_1,\tau_2)\big]\,,\quad \Sigma(\tau_1,\tau_2)=\frac{i}{q}\,\sigma(\tau_1,\tau_2)\,,\quad A(\tau_1,\tau_2)=\frac1{q}\,a(\tau_1,\tau_2)
\end{equation}
where all new bilocal fields $g$, $\sigma$ and $a$ are real, but $g$ is symmetric while $\sigma$ and $a$ are antisymmetric
These new fields will be kept finite in the limit $q\to\infty$. Inserting \eqref{fieldredefs} in \eqref{collect} and solving for the auxilliary field $\sigma$ via its own equation of motion gives $\hat\sigma =\hat a-i \,\hat K_1\,\hat\gamma\,\hat K_1+\mathcal{O}(\tfrac1{q})$, where $\gamma(\tau_1,\tau_2):=G_0(\tau_1,\tau_2)\,g(\tau_1,\tau_2)$. This leads to
\begin{eqnarray}\label{collect2}
&&  I[A,G,\Sigma\,;N,q]-I[A,0,0\,;N,q]\nonumber\\&&=\,\frac{N}{q^2}\,\Bigg(-\text{Tr}\Big[\,i\hat A\,\hat\gamma+\frac12\,\hat K_1\,\hat\gamma\,\hat K_1\,\hat\gamma \,\Big]
    +J^2\int d\tau_1 \, d\tau_2\,\underbrace{G_0(\tau_1,\tau_2)^q}_{=1}\;e^{g(\tau_1,\tau_2)}+\mathcal{O}\left(\tfrac1{q}\right)\Bigg)\,.
\end{eqnarray}
The front factor $\frac1{e^2}:=\frac{N}{q^2}$ will be kept finite in the double scaling limit. 

Let us finally define the real symmetric field  
$h(\tau_1,\tau_2):=\text{sgn}(\tau_1,\tau_2)\,a(\tau_1,\tau_2)$
to conclude that
\begin{eqnarray}\label{Liouvilleact}
S[h,g\,;J,e]&:=& \lim\limits_{q\to\infty}\Big(I[A,G,\Sigma\,;N,q]-I[A,0,0\,;N,q]\Big)\nonumber\\&=&\frac{1}{e^2}\int d\tau_1 \, d\tau_2\,\Big(\,\frac12\,\partial_{\tau_1}g(\tau_1,\tau_2)\partial_{\tau_2}g(\tau_1,\tau_2)+J^2e^{g(\tau_1,\tau_2)}+h(\tau_1,\tau_2)\,g(\tau_1,\tau_2)\Big)
\end{eqnarray}
up to (ultra)local terms (i.e. with a single $\tau$ integral).
The above action has the form of two-dimensional Liouville action in the presence of a source, where $\tau_1$ and $\tau_2$ play the role of lightcone coordinates. Let us stress that the dynamical field is symmetric, $g(\tau_1,\tau_2)=g(\tau_2,\tau_1)$, in contradistinction with usual Liouville theory. In particular, this symmetry condition on the scalar field breaks by half the conformal symmetries of two-dimensional Liouville theory, in agreement with the conformal symmetry of the original one-dimensional theory. 
It is in this precise sense that the double scaling limit of SYK model is a Liouville-like theory. Setting $J^2=0$ in the previous derivation, one similarly obtains that the large-$N$ limit of free SYK model is a d'Alembert-like theory.

Adapting the symmetries of Liouville theory analysed in \cite{Zhiber,Kiselev} to the present case (with a symmetry condition on the dynamical field $g$ and with a source $h$ added), one can find the infinitesimal higher-spin symmetries of the sourced Liouville equation (we set $J=1$ for simplicity)
\begin{equation}\label{sourcedLiouville}
    \partial_{\tau_1}\partial_{\tau_2}g(\tau_1,\tau_2)- \exp\big[\,g(\tau_1,\tau_2)\,\big]\,=\,h(\tau_1,\tau_2)\,,
\end{equation}
with the symmetry conditions $g(\tau_1,\tau_2)=g(\tau_2,\tau_1)$ and $h(\tau_1,\tau_2)=h(\tau_2,\tau_1)$.
The higher-spin symmetries of the sourceless Liouville equation $\partial_{\tau_1}\partial_{\tau_2}\,g=\exp\,g$ read explicitly
\begin{equation}
    \delta g=\big(\partial_{\tau_1}+g\big)\varepsilon\big(\tau_1\,,T_1\,,\partial_{\tau_1}T_1\,,\ldots\big)+\big(\partial_{\tau_2}+g\big)\varepsilon\big(\tau_2\,,T_2\,,\partial_{\tau_2}T_2\,,\ldots\big)\,,
\end{equation}
where $\varepsilon(\tau_i\,,T_i\,,\partial_{\tau_i}T_i\,,\ldots)$ is a local function of the left/right on-shell energy-momentum tensor
\begin{equation}
    T_i\,:=\,\partial_{\tau_i}^2g-\tfrac12(\partial_{\tau_i}g)^2\,,\qquad (i=1,2)\,.
\end{equation}
This is easy to check by making use of the operatorial identities
\begin{eqnarray}
\left(\partial_{\tau_1}\partial_{\tau_2}-e^{g}\right)\circ\big(\partial_{\tau_1}+g\big)&=&\partial_{\tau_1}\circ\Big(\big(\partial_{\tau_1}+g\big)\circ\partial_{\tau_2}+\partial_{\tau_1}\partial_{\tau_2}g-e^{g}\Big)\\
\left(\partial_{\tau_1}\partial_{\tau_2}-e^{g}\right)\circ\big(\partial_{\tau_2}+g\big)&=&\partial_{\tau_2}\circ\Big(\big(\partial_{\tau_2}+g\big)\circ\partial_{\tau_1}+\partial_{\tau_1}\partial_{\tau_2}g-e^{g}\Big)
\end{eqnarray}
and the property that, for solutions of the (sourceless) Liouville equation, 
\begin{equation}
    \partial_{\tau_2}T_1=0=\partial_{\tau_1}T_2\,.
\end{equation}
The latter equations do not hold for the equation \eqref{sourcedLiouville} when $h\neq 0$. In this case, the source must transform as
\begin{eqnarray}
    \delta h&=&\partial_{\tau_1}\left[\varepsilon(\tau_1\,,T_1\,,\ldots)\, h+\big(\partial_{\tau_1}+g\big)\left(\frac{\partial\varepsilon(\tau_1\,,T_1\,,\partial_{\tau_1}T_1\,,\ldots)}{\partial T_1}\,\big(\partial_{\tau_1}h-h\,\partial_{\tau_1}g\big)+\cdots\right)\right]\\
    &&\quad\,+\,\partial_{\tau_2}\left[\varepsilon(\tau_2\,,T_2\,,\ldots)\, h+\big(\partial_{\tau_2}+g\big)\left(\frac{\partial\varepsilon(\tau_2\,,T_2\,,\partial_{\tau_2}T_2\,,\ldots)}{\partial T_2}\,\big(\partial_{\tau_2}h-h\,\partial_{\tau_2}g\big)+\cdots\right)\right]\,.\nonumber
\end{eqnarray}
Similarly, one can also write the off-shell symmetries (i.e. at the level of the action) but they are slightly more cumbersome to describe, so we do not explicit them.

\section{Traces on the algebra $\mathcal{A}[\lambda,\nu]$}
\label{app:trace}

In this Appendix, we classify traces on the two-parameter family of associative algebras $\mathcal{A}[\lambda,\nu]$, which we denote  simply by $\mathcal A$ here. 

In purely algebraic terms, classifying traces on $\mathcal{A}$ is equivalent to computing the zeroth Hochschild cohomology group $HH^0(\mathcal{A},\mathcal{A}^\ast)$. 
The elements of this group are defined as linear functionals $\mathrm{tr}: 
\mathcal{A}\rightarrow \mathbb{C}$ that vanish on commutators, meaning  that $\mathrm{tr}(a)=0$ for all $a\in [\mathcal{A}, \mathcal{A}]$. Equivalently, the traces on the algebra $\mathcal{A}$ form a linear space that is dual to the quotient $\mathcal{A}/[\mathcal{A},\mathcal{A}]$. To evaluate the dimension of the quotient space and its dual, we first need to evaluate the size of the commutator subspace $[\mathcal{A},\mathcal{A}]\subset \mathcal{A}$. Recall that $\mathcal{A}$ is a finitely-generated unital associative algebra defined by the relations \eqref{kkk}-\eqref{hee} on the four generators $e_+$, $h$, $e_-$, $k$, or, equivalently:
\begin{equation}\label{heeapp}
\begin{array}{c}
[h,e_{\pm}]=\pm e_{\pm}\,,\qquad hk=kh\,,\qquad e_\pm k=-ke_\pm\,,\qquad k^2=1\,,\\ [5mm] 
\displaystyle e_+e_- =\frac12\Big[\frac14(\lambda^2 -1)-h^2+\frac{\nu}{2}k-\frac{\nu^2}{4}+(1+\nu k)h\Big]\,,\\[5mm]
\displaystyle e_-e_+ = \frac12\Big[\frac14(\lambda^2 -1)-h^2+\frac{\nu}{2}k-\frac{\nu^2}{4} -(1+\nu k)h\Big]\,.
\end{array}
\end{equation}
We assume that the algebra $\mathcal{A}$ satisfies the Poincar\'e--Birkhoff--Witt (PBW) property. Under this assumption, we can identify a convenient basis of monomials: 
$$ E^\pm_{nm}:=h^ne_\pm^m\,,\qquad \bar E_{nm}^\pm:= k h^ne_\pm^m\,, \qquad n,m=0,1,2,\ldots$$
When multiplying two basis elements, the following formula is particularly useful:
\begin{equation}
    e_{\pm}^mh^n=(h\mp m)^ne^m_\pm\,.
\end{equation}
The commutation relations 
$$
[h,E_{nm}^+]=mE_{nm}^+\,,\qquad [h,E^-_{nm}]=(-1)^m mE_{nm}^-\,, 
$$
$$[h,\bar E_{nm}^+]=m\bar E_{nm}^+\,, \qquad [h,\bar E_{nm}^-]=(-1)^m m\bar E_{nm}^-\,,
$$
imply that $\mathrm{tr}(E^\pm_{nm})=\mathrm{tr}(\bar E^\pm_{nm})=0$ whenever $m>0$. It remains to determine the value of a trace on the subspace $V\subset \mathcal{A}$ spanned by the powers $h^n$ and $kh^n$. By definition, the trace must vanish on the intersection $W=V\cap [\mathcal{A}, \mathcal{A}]$. The subspace $W\subset V$ is generated by the commutators 
\begin{equation}\label{basis}
    [E^+_{np}, E^-_{mp}]=[\bar E^+_{np}, \bar E^-_{mp}]\,,\qquad   [\bar E^+_{np},  E^-_{mp}]=(-1)^p[E^+_{np}, \bar E^-_{mp}]
\end{equation}
for various $m$, $n$, and $p$. This generating set  is highly redundant, meaning that there exist a number of linear relations among the commutators. To isolate a convenient basis in $W$ we first find that
\begin{equation}
    [e_+,h^ne_+^p]=f_{n}(h)e_+^{p+1 }\,,  \qquad [ke_+, h^ne_+^{2l}]=kf_{n}(h)e_+^{2l+1}\,, \qquad  [ke_+, h^ne_+^{2l+1}]=k\bar f_{n}(h)e_+^{2l+2}\,,
\end{equation}
for some 
\begin{equation}
    \qquad f_{n}(h)=-nh^{n-1}+ \mathcal{O}(h^{n-2})\,,\qquad \bar f_n(h)=2h^{n}+\mathcal{O}(h^{n-1})\,.
    \end{equation}
Considered individually, the polynomials $\{f_n\}_{n=1}^\infty$ and $\{k \bar f_n\}_{n=0}^\infty$ are linearly independent and generate bases in $V$. Therefore, we can write
\begin{equation}
E^+_{mp}= [e_+, g_{m}(h)e^{p-1}_+ ]\,,\qquad \bar E^+_{m, 2l+1}= [ke_+, g_{m}(h)e^{2l}_+ ]\,, \qquad \bar E^+_{m, 2l+2}= [ke_+, \bar g_{m}(h)e^{2l+1}_+ ]
\end{equation}
for $p \geqslant 1$, $l\geqslant 0$, and some polynomials  $\bar g_m(h)$ and $g_{m}(h)$ of degree $m$ and $m+1$, respectively. Using the Jacobi identity, we then find
\begin{equation}
    [E^+_{mp},E^-_{np}]=\big[\,[e_+, g_{m}(h)e^{p-1}_+]\,, E^-_{np}\, \big]=\big[\,e_+\,, [g_{m}(h)e^{p-1}_+ , E^-_{np}]\,\big]-\big[\,g_{m}(h)e^{p-1}_+, [e_+,E^-_{np}]\,\big] 
\end{equation}
$$
=[\,e_+\,, u_{mnp}(h,k)e_-\,]+[\,g_{m}(h)e^{p-1}_+,v_{np}(h,k)e_-^{p-1}\,]\,,
$$
$$
[\bar E^+_{m,2l+1},E^-_{n,2l+1}]=\big[\,[ke_+, g_{m}(h)e^{2l}_+]\,,E^-_{n,2l+1}\,\big]=\big[\,ke_+, [g_{m}(h)e^{2l}_+,E^-_{n,2l+1}]\,\big]-\big[\,g_{m}(h)e^{2l}_+\,,[ke_+,E^-_{n,2l+1}]\,\big]
$$
$$
=\big[ke_+\,,u_{m,n,2l+1}(h,k)e_-]+[g_{m}(h)e^{2l}_+, kv'_{n,2l+1}(h,k)e_-^{2l}]\,,
$$
$$
[\bar E^+_{m,2l},E^-_{n,2l}]=\big[\,[ke_+, \bar g_{m}(h)e^{2l-1}_+]\,,E^-_{n,2l}\,\big]=\big[\,ke_+\,, [\bar g_{m}(h)e^{2l-1}_+,E^-_{n,2l}]\,\big]-\big[\,\bar g_{m}(h)e^{2l-1}_+\,,[ke_+,E^-_{n,2l}]\,\big]
$$
$$
=[\,ke_+\,,u'_{mnl}(h,k)e_-\,]+[\,\bar g_{m}(h,k)e^{2l-1}_+, kv'_{n,2l}(h,k)e_-^{2l-1}\,]\,,
$$
where the $u$'s and $v$'s are some polynomials in $h$ and $k$. 
Each of the commutators $[E_{mp}^+, E^-_{np}]$ and $[\bar E^+_{mp}, E^-_{np}]$ is thus expressed as a linear combination of the commutators $[E_{sq}^+, E^-_{tq}]$ and $[\bar E^+_{sq}, E^-_{tq}]$ with $q=1,2,\ldots, p-1$. By induction on $p$, we conclude that the space $W$ is linearly generated 
by the set of  commutators 
\begin{equation}
    [E_{m1}^+, E^-_{n1}]\,,\qquad [\bar E_{m1}^+, E^-_{n1}]
\end{equation}
for $m,n=0,1,2,\dots$ These commutators are still linearly dependent. Indeed, 
\begin{align}
[E^+_{m1}, E^-_{n1}]&=[h^me_+,h^ne_-]  = h^{m-1}e_+h^{n+1}e_- +h^{m-1}[h,e_+]h^{n}e_--h^{n+1}e_-h^{m-1}e_+      -h^{n}[e_-,h]h^{m-1}e_+\notag\\
     &= h^{m-1}e_+h^{n+1}e_-+h^{m-1}e_+h^{n}e_--h^{n+1}e_-h^{m-1}e_+  -h^{n}e_-h^{m-1}e_+       \notag\\
     &=[h^{m-1}e_+, h^{n+1}e_-]+[h^{m-1}e_+,h^{n}e_- ]  =[E_{m-1,1}^+,E_{n+1,1}^-] +[E^+_{m-1,1},E^-_{n,1}]\notag
\end{align}     
and, similarly,
\begin{equation}
         [\bar E^+_{m1},E^-_{n1}]=[\bar E^+_{m-1,1}, E^-_{n+1,1}]+[\bar E^+_{m-1,1},E^-_{n1}]\,.
\end{equation}
Again, by induction on $m$, we see that the space $W$ is spanned by the commutators
\begin{equation}\label{a10}
   [E^+_{01}, E^-_{n1}] =[e_+,h^ne_-]=[e_+,h^n]e_-+h^n[e_+,e_-]=\big((h-1)^n-h^n\big)e_+e_-+(1+\nu k)h^{n+1}
\end{equation}
$$
=(1+\nu k)h^{n+1}+\frac12\Big((1+\nu k)h-h^2+\frac14(\lambda^2 -1)+\frac\nu2 k-\frac{\nu^2}4\Big)\Big((h-1)^n-h^n\Big)
$$
$$
=\Big(\frac{n+2}2+\nu k\Big) h^{n+1}+\mathcal{O}(h^n)
$$
and
\begin{equation}\label{a11}
[\bar E_{01}^+, E^-_{n1}]=[ke_+,h^n e_-]=[ke_+,h^n]e_-+h^n[ke_+,e_-]=k\big((h-1)^n-h^n\big)e_+e_-+kh^{n}(e_+e_-+e_-e_+)
\end{equation}
$$
=\frac k2\Big((1+\nu k)h-h^2+\frac14(\lambda^2 -1)+\frac\nu2 k-\frac{\nu^2}4\Big)\Big((h-1)^n-h^n\Big)
-kh^n\Big(h^2-\frac14(\lambda^2 -1)-\frac{\nu}{2}k+\frac{\nu^2}{4}\Big)
$$
$$
=-kh^{n+2}+O(h^{n+1})
$$
for $n=0,1,2,\ldots$ These commutators are now linearly independent and form a basis for $W$. It is evident that the elements of the quotient space $V/W$ are represented by linear combinations of the three  monomials $1$, $k$ and $kh$. Hence, $\dim V/W=3$, or equivalently, $HH^0(\mathcal{A},\mathcal{A}^\ast)\simeq\mathbb{C}^3$. By prescribing arbitrary values to $\mathrm{tr}(1)$, $\mathrm{tr}(k)$, and $\mathrm{tr}(kh)$, we can uniquely determine the traces of all other elements of $\mathcal{A}$. For instance, setting $n=0,1$ in Eqs. (\ref{a10}) and (\ref{a11}), we obtain
\begin{equation}
    \mathrm{tr}(h)=-\nu\mathrm{tr}(kh)\,,\qquad \mathrm{tr}(k h^2)=\frac{\nu}2\mathrm{tr}(1)+\Big[\frac14(\lambda^2 -1)-\frac{\nu^2}{4}\Big]\mathrm{tr}(k)\,,
\end{equation}
$$
\mathrm{tr}(h^2)=\frac13\Big[\frac14(\lambda^2 -1)+\frac{\nu^2}2\Big]\mathrm{tr}(1)-\nu\Big[\frac14(\lambda^2 -1)-\frac16-\frac{\nu^2}{4}\Big]\mathrm{tr}(k)\,.
$$

The construction of the action (\ref{EHS}) requires a trace that defines a non-degenerate inner product $\langle a,b\rangle:=\mathrm{tr}(ab)$ 
on the algebra $\mathcal{A}$.  For $\nu=0$, such a trace is known to exist and is characterized  by the following conditions:  
\begin{equation}\label{ntr}
    \mathrm{tr}(1)=1\,,\qquad \mathrm{tr}(k)=0\,,\qquad \mathrm{tr}(k h)=0\,.
    \end{equation}
In fact, this trace is completely determined by a non-degenerate trace on the subalgebra $\mathrm{Mat}[\lambda]\subset \mathcal{A}[\lambda]$. 
 By continuity, this trace remains non-degenerate on the deformation $\mathcal{A}[\lambda,\nu]$ for $\nu$ small enough, which is sufficient for the perturbative construction  of  
 the action (\ref{EHS}).

\footnotesize
\providecommand{\href}[2]{#2}\begingroup\raggedright\endgroup

\end{document}